\documentclass[fleqn,usenatbib]{mnras}

\usepackage[T1]{fontenc}
\usepackage{ae,aecompl}

\usepackage{graphicx}   
\usepackage{amsmath}    
\usepackage{amssymb}    
\usepackage{pdflscape}
\usepackage{multirow}
\usepackage{newtxtext,newtxmath}

\newcommand{\kep}{\textit{Kepler} }
\newcommand{\ktwo}{\textit{K2} }

\newcommand{\todcor}{\textsc{todcor} }
\newcommand{\jkt}{\textsc{jktebop} }
\newcommand{\jktabs}{\textsc{jktabsdim} }
\newcommand{\vfit}{\textsc{v2fit} }
\newcommand{\ms}{m~s$^{-1}$}
\newcommand{\fdb}{\textsc{fd3} }
\newcommand{\ispec}{\textsc{ispec} }

\title[ASAS systems in {\it K2}]{Orbital and physical parameters of eclipsing binaries from the ASAS catalogue - XII. A sample of systems with {\it K2} photometry}

\author[K. G. He{\l}miniak et al.]
{K. G. He{\l}miniak,$^{1}$\thanks{E-mail: xysiek@ncac.torun.pl (KGH)}
A. Moharana,$^{1}$
T. Pawar,$^{1}$
N. Ukita,$^{2,3}$
P. Sybilski,$^{1}$
\newauthor
N. Espinoza,$^{4}$
E. Kambe,$^{5}$
M. Ratajczak,$^{6}$
A. Jord\'{a}n,$^{7,8}$
H. Maehara,$^{2}$
\newauthor
R. Brahm,$^{7,8}$
S. K. Koz{\l}owski,$^{9}$
and M. Konacki.$^{9}$
\\
$^{1}$Nicolaus Copernicus Astronomical Center, Polish Academy of Sciences, ul. Rabia\'{n}ska 8, 87-100 Toru\'{n}, Poland\\
$^{2}$Okayama Astrophysical Observatory, National Astronomical Observatory of Japan, 3037-5 Honjo, Kamogata, Asakuchi,\\ Okayama 719-0232, Japan\\
$^{3}$The Graduate University for Advanced Studies, 2-21-1 Osawa, Mitaka, Tokyo 181-8588, Japan\\
$^{4}$Space Telescope Science Institute, 3700 San Martin Dr., Baltimore 21218, MD, USA\\
$^{5}$Subaru Telescope, National Astronomical Observatory of Japan, 650 North Aohoku Place, Hilo, HI 96720, USA\\
$^{6}$Warsaw University Astronomical Observatory, Al. Ujazdowskie 4, 00-478 Warszawa, Poland \\
$^{7}$Facultad de Ingenier\'ia y Ciencias, Universidad Adolfo Ib\'a\~nez, Av.\ Diagonal las Torres 2640, Pe\~nalol\'en, Santiago, Chile\\
$^{8}$Millennium Institute for Astrophysics, Santiago, Chile\\
$^{9}$Institute of Automatic Control and Robotics, Pozna\'{n} University of Technology, Piotrowo 3A, 61-131 Pozna\'{n}, Poland\\
$^{10}$Nicolaus Copernicus Astronomical Center, Polish Academy of Sciences, ul. Bartycka 18, 00-716 Warszawa, Poland
}

\date{Accepted XXX. Received YYY; in original form ZZZ}

\pubyear{2020}

\begin{document}
\label{firstpage}
\pagerange{\pageref{firstpage}--\pageref{lastpage}}
\maketitle

\begin{abstract}
We present results of the analysis of light and radial velocity (RV) curves of eight detached eclipsing binaries observed by the All-Sky Automated Survey, which we have followed up with high-resolution spectroscopy, and were later observed by the \kep satellite as part of the \ktwo mission. 
The RV measurements came from spectra obtained with OAO-188/HIDES, MPG-2.2m/FEROS, SMARTS 1.5m/CHIRON, Euler/CORALIE, ESO-3.6m/HARPS, and OHP-1.93/ELODIE instruments. 
The \ktwo time-series photometry was analysed with the \jkt code, with out-of-eclipse modulations of different origin taken into account. Individual component spectra were retrieved with the \fdb code, and analysed with the code \ispec in order to determine effective temperatures and metallicities.
Absolute values of masses, radii, and other stellar parameters are calculated, as well as ages, found through isochrone fitting. 
For five systems such analysis has been done for the first time. 
The presented sample consists of a variety of stars, from low-mass dwarfs, through G and F-type Main Sequence objects, to evolved active sub-giants, one of which is found to be crossing the Hertzsprung gap. One target may contain a $\gamma$ Dor-type pulsator, two more are parts of higher-order multiples, and spectra of their tertiaries were also retrieved and used to constrain the properties of these systems. 

\end{abstract}




\section{Introduction}

Double-lined detached eclipsing binaries (DEBs) are among the most useful objects in 
astrophysics. Their photometric and spectroscopic observations allow one to determine 
basic stellar parameters and carry out a wide range of tests of stellar structure, 
evolution and dynamics. They play important role for such branches of astronomy as 
galaxy structure and evolution, stellar populations, formation and evolution of 
exoplanets or tidal interactions. In order to be considered useful, a DEB needs to have 
parameters of its components (i.e. mass and radius) derived with the accuracy of $<$3\% 
\citep{cla08,las02}, and even down to $\sim$0.1\% to challenge the modern models of 
stellar evolution \citep{val17}.
A surprisingly low number of DEBs have their absolute masses and radii derived accurately 
enough. In an up-to-date DEBCat catalog
\citep{sou15}\footnote{\url{https://www.astro.keele.ac.uk/~jkt/debcat/}},
there are only 271 systems listed (at the moment of writing this article). In a groundbreaking 
review, \citet{tor10} point out the lack of well-characterized stars at the high- and 
low-mass end of the main sequence, or on very early or late stages of evolution. In the 
same time they notice that evolutionary models often fail to reproduce the observed 
values of parameters for a variety of objects, due to improper treatment of the convection, 
overshooting or magnetic fields for instance. Future models will attempt to fix these 
discrepancies, but they can only be verified with a sufficient number of high-quality data. 
Such data can only come from observations of many objects of a particular class. Meanwhile,
except for F, G and early K-type main sequence (0.8-2.0~M$_\odot$) stars, we see a lack of 
sufficient number of well-studied objects.

As a response to Torres' call, in 2011 we started a large spectroscopic 
programme, aimed for an overall characterization of several hundred DEBs, including 
numerous new cases of stars occupying the under-populated regions in both southern 
\citep[see other papers of this series, e.g.:][]{hel14,hel15,hel19,rat13,cor15}
and northern hemispheres \citep[e.g.][]{hel19b}. This survey provided
thousands of high-resolution spectra and precise radial velocity (RV) measurements,
which need to be supplemented with high-precision photometry.
The most precise are from space-based observatories, such as  
{\it Kepler}, or TESS. We exploited the \kep main field \citep{hel16,hel19b}, 
characterizing a majority of bright ($V<12$~mag) and not previously studied DEBs, 
and the ongoing TESS observations are already providing data for many of our targets 
\citep{hel19c,mar20,rat20}. 
In this work we focus on those systems from our survey, that fell into the field
of view of the \kep satellite during its \ktwo mission.

The paper is organized as follows. 
In Section~\ref{sec_targets} we present our sample of DEBs; 
in Sect.~\ref{sec_data} we present the spectroscopic and photometric data sets; 
Sect.~\ref{sec_ana} describes the methodology of data analysis used in this work;
Sect.~\ref{sec_res} presents the results, including estimates of ages; 
and finally Sect.~\ref{sec_conc} concludes our findings.

\begin{table*}
    \centering
    \caption{Basic information about the presented systems, as listed in {\it Simbad}. 
    Positions, proper motions and distances come from the {\it Gaia} Early Data Release 3 (GEDR3) catalogue
    \citep{gai21}. The $K$-band magnitudes are from 2MASS \citep{cut03}.}
    \label{tab_lit}
    \begin{tabular}{lllccccccc}
    \hline \hline
    ASAS ID & EPIC ID &  Other ID & RA & DEC & $\mu_{\rm RA}$ & $\mu_{\rm DEC}$  & $V$ & $K$ & $d_{GEDR3}$ \\
            &      &    & ($^\circ$) & ($^\circ$) & (mas yr$^{-1}$) & (mas yr$^{-1}$) & (mag) & (mag) & (pc) \\
    \hline
045021$+$2300.4 & 247605441 & HD 284753   &\ 72.585989 & +23.005982 &  1.4(4)  & -40.9(3) & 10.26 &\ 8.176 & 100(3)  \\ 
060505$+$2032.1 & 202073040 &TYC 1321-16-1&\ 91.270401 & +20.536897 &  1.64(2) & -4.72(2) & 11.93 & 10.814 & 1095(25)\\ 
083730$+$2333.7 & 212173112 & RU Cnc	  & 129.375422 & +23.561564 &-21.62(2) & -0.81(1) & 10.20 &\ 8.015 & 408(3)  \\ 
085002$+$1752.5 & 211839462 & BD+18 2050A & 132.505874 & +17.874540 & -9.06(2) & -4.20(1) & 10.15 &\ 9.145 & 522(4)  \\ 
111245$+$0020.9 & 201488365 & FM Leo	  & 168.187464 &\ +0.347869 &-96.73(3) &-32.24(2) &\ 8.45 &\ 7.211 & 146.4(6) \\ 
163903$-$2847.2 & 202674012 & HD 149946   & 249.764233 &$-$28.787116& 15.49(2) &-11.94(2) &\ 9.85 &\ 8.326 & 267.3(1.4)\\
171750$-$1915.3 & 234440875 & BD-19 4582  & 259.456832 &$-$19.254559& -6.75(2) &-10.98(1) & 10.99 &\ 9.178 & 368(2)   \\ 
231922$-$0852.2 & 246024234 & HD 219869   & 349.842294 &\ $-$8.870378&52.36(2) & 10.63(2) & 10.24 &\ 8.990 & 289.6(1.6)\\
    \hline
    \end{tabular}
\end{table*}

\section{Targets}\label{sec_targets}

All presented DEBs were observed and/or discovered by the All-Sky Automated Survey 
\citep[ASAS;][]{poj02}, including its northern counterpart (ASAS-N). With one exception, 
they are listed in the ASAS Catalog of Variable Stars (ACVS). They were included into target 
list of the previously mentioned, large spectroscopic survey, which we conducted in years 
2011-2018 on a number of telescopes and spectrographs, and which focused on relatively bright ($V<12$~mag)
detached binaries of spectral types F to M (observed colour $V-K > 1.1$ mag).
Spectroscopic observations were done independently from and started 
usually before the \ktwo mission. Below we briefly describe each of our targets 
(ordered by increasing right ascension):

\begin{itemize}
    \item {\it ASAS J045021+2300.4 = EPIC 247605441, HD 284753,  BD+22 760} (hereafter: A-045):
This system was first recognized as an eclipsing variable star from the ASAS-N photometry 
by \citet{kir13}, and is the only case from this work not listed in the ACVS. Nevertheless, 
it is close to the northern edge of the field of view of the southern ASAS station 
($\delta<+28^\circ$), so its pre-2009 photometry is also available. It has been associated 
with a ROSAT Bright Source Catalogue \citep{vog99} object 1RXS~045021.0+230037, which 
suggests its strong activity. No detailed study of this system has been conducted so far.

    \item {\it ASAS J060505+2032.1 = EPIC 202073040, TYC 1321-16-1} (hereafter: A-060):
This system was discovered as a DEB by the ASAS, but the period given in the ACVS ($P\simeq33.371$~d) 
turned out to be incorrect. Later, it was identified in \ktwo photometry first by \citet{arm15}, 
and shortly after by \citet{lac15}. The correct orbital period ($P\simeq2.121$~d) was found only 
in the latter work, while \citet{arm15} gave a value close to $P/2$. No detailed study of this system 
has been conducted so far.
    
    \item {\it ASAS J083730+2333.7 = EPIC 212173112, RU Cnc, BD+24 1959, HIP 42303} (hereafter: A-083):
This star was discovered as a variable in 1908 by L.~Ceraski, 
and designated at that time as 3.1911~Cancri \citep{cer11}. It is first mentioned as an eclipsing binary
by \citet{sha13}. Numerous studies were published since then, including two with RV curves: \citet{pop90},
and \citet{imb02}. \citet{pop90} also performed a light curve analysis, and estimated absolute values of
masses and radii. Majority of estimates of the effective temperature that are available in the literature,
were done under the assumption of the target being a single star, therefore they are not reliable. 
Most recently, \citet{cok19} analysed the \ktwo photometric data and the two historical RV curves.
In this work, we calculated the RVs from our own, as well as unpublished archival spectra, 
and used a different approach to the light curve 
analysis than \citet{cok19}.

    \item {\it ASAS J085002+1752.5 = EPIC 211839462/-30, BD+18 2050 AB, ADS 7030 AB} (hereafter: A-085): 
This DEB, discovered as an eclipsing variable by ASAS, and independently identified 
by the KELT survey \citep{pep08}, is a part of a visual pair ADS~7030~AB (a.k.a. 
WDS~08500+1752AB, CCDM~J08500+1752AB, KU~33), whose components differ in brightness by about 
0.3~mag. The corresponding EPIC designations are 211839462 for the brighter, western star A, 
and 211839430 for the fainter component B to the East. The component B is currently incorrectly 
noted as a DEB in {\it Simbad} and in the Washington Double Star Catalogue \citep[WDS;][]{mas01}, 
probably because of the original ASAS identification. However, our spectroscopic observations 
revealed a single-lined spectrum of the component B, and a double-lined and rotationally broadened 
spectrum of component A, with RV changes clearly seen. This has been confirmed by the analysis of 
\ktwo data, when deeper eclipses were identified on the position of EPIC...62 rather
than on ...30 \citep{bar16}. 
Atmospheric parameters, such as effective temperature $T_{\rm eff}$, logarithm of gravity $\log(g)$ 
or metallicity [M/H] for the component B can be found in the literature \citep{hub16,ho17,tin18,gai18}, 
all pointing to a red giant with $T_{\rm eff}\simeq5000$~K, but not very consistent in terms of 
other values. \citet{lee15} used the ASAS light curve and the MECI \citep{dev08} code to find the 
most probable masses and age of the eclipsing binary, but only got limits for these parameters.

    \item {\it ASAS J111245+0020.9 = EPIC 201488365, FM Leo, HD 97422, BD+01 253, HIP 54766} (hereafter: A-111):
This DEB was discovered by the \textit{Hipparcos} mission, and found its place in the 74th Special 
Name-List of Variable Stars \citep{kaz99}. The first RV and light curve solutions were presented 
by \citet{rat10}, but the quality of the photometric data was far from optimal. More recently, 
\citet{max18} combined the \ktwo data with RVs from \citet{rat10}, and obtained very precise results.
Another set of RV measurements, including large number of spectra taken during eclipses, was combined 
with the ASAS light curve and used to model the system by \citet{syb18}, who also showed the spin-orbit 
alignment of the two components. Finally, \citet{gra21} used new HARPS spectra, together with the 
\ktwo photometry, and presented a comprehensive analysis, including spectra decomposition, 
atmospheric parameters, and eclipse timing variations.
In this work, we use  our own RVs calculated from: (1) the out-of-eclipse spectra from 
\citet{syb18}\footnote{FM Leo is an exception in the sample, since after the work 
of \citet{rat10} we did not intend to include it into our large spectroscopic survey. We changed this 
after P. Sybilski realized that this star was very well suited for studying the Rossiter-McLaughlin 
effect.}, (2) the HARPS spectra used by \citet{gra21}, and (3) publicly available FEROS spectra,
not used in any publication so far. We also present our own approach to the \ktwo data,
independently from \citet{max18} and \citet{gra21}.

\item {\it ASAS J163903-2847.2 = EPIC 202674012, HD 149946} (hereafter: A-163):
This system was first identified as a DEB by ASAS. Recently, the \ktwo light curve was analysed by 
\citet{max18} together with four publicly available FEROS spectra from our spectroscopic survey, leading 
to a preliminary solution for this object. Later, in a short Research Note \citep{hel18}, we presented
a more precise orbital solution from 15 high-resolution spectra, which we combined with LC-based
parameters from \citet{max18}. Recently, \citet{hoy20} presented their study, in which they
used CORALIE, CHIRON and one FEROS RV measurements from \citet{hel18}, their own measurements 
for three archival FEROS and three HARPS spectra, own light curve modelling, and atmospheric 
parameters obtained from disentangled component spectra.
In this work, we repeat the orbital analysis with the addition of ten HARPS spectra
from the ESO archive, and explain the process in more details. Furthermore we present our 
own approach to the \ktwo data, not relying on results of \citet{max18} nor \citet{hoy20}.
    
    \item {\it ASAS J171750-1915.3 = EPIC 234440875, BD-19~4582} (hereafter: A-171): 
Another DEB discovered by ASAS, although the orbital period given in the ACVS is slightly longer than
the true value, which affected the phase-folded light curve. No detailed study of this system has 
been conducted so far.

    \item {\it ASAS J231922-0852.2 = EPIC 246024234, HD~219869, BD-09 6166} (hereafter: A-231): 
This star was first identified as a DEB in the ACVS, and later in data from the STEREO satellite
\citep{wra11}. In both cases, however, the given orbital period is close to half of the true value. 
Two spectroscopic observations were taken by the RAVE survey \citep{kor13,kun17}, and the measured RVs
differ by over 50 k\ms, which indicates a spectroscopic binary. Apart from that, no other study of this 
system has been performed.
\end{itemize}

\section{Data}\label{sec_data}

\subsection{Spectroscopic observations and data reduction}
Spectra of the presented systems come mainly from our own large spectroscopic survey, and were 
taken with four major instruments we used in the project.  

The CHIRON spectrograph \citep{sch12,tok13,par21}, attached to the 1.5-m SMARTS telescope in 
the Cerro Tololo Inter-american Observatory (CTIO, Chile), 
was used in the ``slicer'' and ``fiber'' modes, which provide spectral 
resolution of $R\sim90\,000$ and $\sim$28\,000, respectively. The latter provides
much higher efficiency. This telescope works in service mode only. Spectra were reduced with 
the pipeline developed at Yale University \citep{tok13}. Wavelength 
calibration is based on ThAr lamp exposures taken just before the science 
observation. Barycentric corrections are not applied by the pipeline, thus we
were calculating them ourselves under IRAF\footnote{IRAF is distributed by the 
National Optical Astronomy Observatory (NOAO), which 
is operated by the Association of Universities for Research in Astronomy (AURA) 
under cooperative agreement with the National Science Foundation. 
\url{http://iraf.noao.edu/}} with {\it bcvcor} task. 
For the targets described here we did not use the available iodine (I$_2$) cell. Without 
the I$_2$, the stability of the instrument is estimated to be better than 
15 \ms\ in ``slicer'' mode. For the radial velocity (RV) measurements we used 36 
echelle orders, spanning from 4580 to 6500~\AA\,(limited by the templates we used), 
but the complete spectrum reaches 8760~\AA.

The observations at Okayama Astrophysical Observatory (OAO) 1.88-m telescope in Okayama 
(Japan) with the HIDES spectrograph \citep{izu99,kam13} were conducted in the fibre mode 
with image slicer ($R\sim50\,000$), without I$_2$, and with ThAr lamp frames taken every 
1-2 hours. The spectra are composed of 62 rows covering 4080--7538~\AA, of which we use 30 
(4365--6440~\AA). Detailed description of the observing procedure, data reduction and 
calibrations is presented in \citet{hel16}. The precision reached with our approach is 40-50 \ms.

The CORALIE spectrograph, attached to the 1.2-m Euler telescope in La Silla 
(Chile), works in a simultaneous object-calibration mode, and provides resolution 
of $R\sim70\,000$. Additional ThAr exposures with both fibres are done every 
1-1.5 hours. For this study we used the instrument when it was still
equipped with circular fibres (currently octagonal). Spectra were reduced 
with the dedicated python-based pipeline \citep{jor14,bra17}, which also performs 
barycentric corrections. The pipeline is optimized to derive high-precision radial 
velocities, down to $\sim$5 \ms, and reduces the spectrum to 70 rows spanning 
from 3840 to 6900~\AA. For our purposes, we use only 45 rows (4400--6500~\AA), 
due to the limits of our template spectra and very low signal in the blue part.

Operations at the MPG-2.2m telescope (La Silla, Chile) with the FEROS instrument 
\citep{kau99} look very similar to CORALIE, as the spectrograph also works in 
a simultaneous object-calibration manner, but employs an image slicer, which gives 
$R\sim48\,000$, and the highest efficiency of all the optical instruments we 
used for this study ($>$20\%), thus provides data with the highest SNR. 
Spectra were reduced with a similar pipeline as for CORALIE, capable of providing 
RVs with the precision of 5-8 \ms. Although the original spectral format reaches 
beyond 10\,000~\AA, the output is reduced to 21 rows covering 4115-6519~\AA, of 
which we use 20 (4135--6500~\AA).

A single spectrum of a visual companion to the A-085 system was obtained with
the HDS \citep{nog02} instrument attached to the 8.2-m Subaru telescope, 
located on Maunakea, Hawaii. The observation was taken through a slit $\sim$0.65 wide, 
which resulted in $R\sim55\,000$. The 720-second exposure resulted in SNR$\sim$103. 
A Th-Ar lamp was used for wavelength calibration. 
This spectrum was reduced and calibrated with dedicated IRAF procedures. It was 
only used to evaluate the metallicity of the A-085 system.

Data from these instruments were supplemented with archival spectra from the HARPS
spectrograph, attached to the ESO 3.6-m telescope in La Silla, the FEROS instrument
behind the MPG-2.2m telescope in La Silla, and ELODIE, which was the 
high-resolution instrument of the 1.93-m telescope of the Observatoire de Haute-Provence (OHP).
These data were extracted from the ESO\footnote{\url{http://archive.eso.org/wdb/wdb/adp/phase3_main/form}} 
and ELODIE\footnote{\url{http://atlas.obs-hp.fr/elodie/}} \citep{mou04} archives, respectively.
From FEROS and HARPS observations, we only used spectra reduced and calibrated with the 
local Data Reduction Pipeline (DRP). The HARPS data are available in two modes: 
high-resolution (ECHE; $R\sim115\,000$) and high-efficiency (EGGS; $R\sim80\,000$). 
In case of ELODIE, we retrieved single reconnected spectra, resampled in 
wavelength with a constant step of 0.05~\AA\ (the nominal spectral resolution of the instrument
is $R \sim 42000$), covering the range 4000-6800~\AA.
If archival spectra were used, their number does not exceed 
the number of our own observations, with the exception of A111 = FM~Leo.

Below we summarise spectroscopic data sets for each target separately:
\begin{itemize}
    \item {\it A-045}: The observations were done only with HIDES. They started in December 2014, and lasted till November 2017. A total of 22 spectra were taken that time. Because of the faintness of the secondary, in some cases measurement of its RV was not possible. We therefore have 22 and 17 data points for the primary and secondary, respectively.
    \item {\it A-060}: Observed only with CHIRON in the ``fiber'' mode, mainly in 2014, when eight spectra were recorded. Two additional visits were made in 2019, in order to improve the coverage and verify the existence of a possible long-term RV trend.
    \item {\it A-083}: Observed by us only with HIDES. A total of 11 spectra were taken between December 2014 and December 2015. Five additional unpublished spectra, from March 1998, were found in the ELODIE archive.
    \item {\it A-085}: Observations started on HDS in October 2011, and on CORALIE in March 2012, but because of the confusion on which component of the visual pair is the eclipsing binary, the wrong one was observed. Observations of the correct one started in November 2013, and lasted till May 2014. Four CORALIE and seven CHIRON-fiber spectra were taken that time. One of the CHIRON spectra was taken during the total part of the primary eclipse, and was not used for RV calculations.
    \item {\it A-111}: A large number of CHIRON spectra were taken in the ``slicer'' mode during a dedicated campaign in February and March 2016 \citep{syb18}. From this set, 13 observations were taken outside of the eclipses, and used for disentangling and orbital fit. In this work, we use these disentangled spectra to derive atmospheric parameters. For RV calculations we use the original 13 out-of-eclipse observations, and, additionally, a set of HARPS (ECHE; 12) and FEROS (8) spectra of A-111, both available in the ESO archive. The HARPS data were used in \citet{gra21}, but the FEROS spectra were not. We decided to use them both for RV calculations, but not for disentangling nor spectral analysis. Furthermore, for this target we did not use the RV measurements from \citet{rat10}, which are of lower precision. Their inclusion did not affect the final solution, but led to larger uncertainties. 
    \item {\it A-163}: Five CHIRON spectra in the ``slicer'' mode were taken in 2012, alongside with two CORALIE and two FEROS spectra. Further CORALIE (three) and FEROS (one) visits were made in 2013, and later in 2015 (one for both instruments). These 13 spectra constitute the data used for the orbital solution presented in \citet{hel18}. Ten additional HARPS spectra (seven ECHE and three EGGS), taken in 2018 and 2019, were found in the ESO archive.
    \item {\it A-171}: This target was observed with FEROS six times in 2013, and once in 2015. Six additional CORALIE spectra were taken between July 2013  and June 2016. Finally, two HARPS spectra (ECHE), from May 2018 and April 2019, have been extracted from the ESO archive.
    \item {\it A-231}: We have observed this star with FEROS between June 2012 and June 2015, taking 12 spectra in total. Additionally, three HARPS spectra (EGGS), taken in December 2017, have been extracted from the ESO archive.
\end{itemize}

\subsection{RV measurements}
In our spectroscopic survey we measure radial velocities of components of a binary system using our
own implementation of the \todcor technique \citep{zuc94}. As templates we use synthetic spectra
computed with ATLAS 9 \citep{kur92}, which do not reach wavelengths longer than 6500 \AA. This lowers 
the number of useful echelle orders, but reduces the influence of telluric lines, and cuts off the H$\alpha$
line. We also do not take into account orders with the sodium D lines ($\sim$5900~\AA), as they
are often affected by the interstellar medium. Initially we use templates calculated for effective 
temperatures roughly expected from the spectral type of the binary, but for the final measurements 
and orbital fit, we select templates based on $T_{\rm eff}$ expected for a given mass and radius, 
and rotationally broadened. Uncertainties are calculated with a bootstrap procedure \citep{hel12},
which is sensitive to the SNR of a component, and velocity of rotation.
The \todcor procedure also provides the ratio of component fluxes, that maximizes the value
of the CCF at the position of the resulting RVs. These flux ratios can be later used to rescale 
and renormalise disentangled spectra before their analysis. Their reliability was verified at
several occasions with the aid of spectra taken during a total eclipse, e.g. in \citet{hel15},
\citet{rat20}, or here in the case of A-085.

\subsection{{\it K2} photometric data}

The studied systems were observed by the \kep space telescope during several campaigns of its extended 
\ktwo mission \citep{how14}. The photometric time series data were extracted from the Mikulski Archive 
for Space Telescopes (MAST)\footnote{\url{https://mast.stsci.edu/portal/Mashup/Clients/Mast/Portal.html}}.
We usually took 30-minute long-cadence (LC) data, except for A-111, for which short-cadence (SC) data
also exist. Also available in MAST are SC data for A-083, but we decided not to work on them, and the
reasons behind this decision will be explained later in the text.
The \ktwo data are strongly affected by systematics coming from the pointing of the satellite, and its 
roll angle variations. Several automated algorithms have been developed in order to correct for these 
effects, such as: {\sc everest} \citep{lug16,lug18}, {\sc k2sff} \citep{van14}, or {\sc polar} 
\citep{bar16}. They produce light curves designated as High-Level Science Products (HLSP), available
for download from the MAST as calibration level 4 data. Apart from them, the Archive contains flux 
measurements with lower calibration level 2, which include Simple Aperture Photometry (SAP) data,
and measurements corrected with the Presearch Data Conditioning (PDC) method \citep{stu12}. In this 
paper we will refer to the latter as {\sc k2pdc}.

Not all HLSP products are available for the studied systems. Also, each algorithm tackles 
short- and long-term systematic trends in different ways, therefore produces different light 
curves. In some occasions this can affect the intrinsic out-of-eclipse variations, or 
even depths and shapes of the eclipses. For these reasons we have collected all available data
(calibration level 2 and 4), and have inspected them for each system and campaign individually. 
We mainly took into account the stability of the shape and depth of eclipses, as well as 
the out-of-eclipse variations and number of outliers. Sometimes modelling was done on two 
different products, and results of the one that produced lower parameter errors and $rms$ 
of the fit were adopted. 

\begin{table}
    \centering
    \caption{Summary of \ktwo data used in this study}
    \label{tab_lcsum}
    \begin{tabular}{ccccl}
    \hline \hline
    EPIC ID & Cam. & Adopted &  No. of data & Notes \\
    \hline
    247605441 & 13 & {\sc everest} & 3939 & 5 parts \\
    202073040 &  0 & {\sc everest} & 1528 &  \\
    212173112 &  5 & {\sc everest} & 3577 & 3 parts \\
        ''    & 18 & {\sc everest} & 2329 & 2 parts \\
    211839462 &  5 & {\sc everest} & 3596 & 3 parts \\
        ''    & 16 & {\sc everest} & 3857 & 5 parts \\
        ''    & 18 & {\sc everest} & 2456 & 2 parts \\
    201488365 &  1 & {\sc everest} & 3513 & 5 parts, SC \\
    202674012 &  2 & {\sc everest} & 3247 &  \\
    234440875 & 11 & {\sc k2pdc}   &  745 & C11a only \\
    246024234 & 12 & {\sc everest} & 3383 & 3 parts \\
        ''    & 19 & {\sc k2sff}   &  339 & Single period \\
    \hline
    \end{tabular}
\end{table}

Table~\ref{tab_lcsum} summarizes the \ktwo data adopted for each system and campaign. 
Figure~\ref{fig_lcsum} in the Appendix shows all the adopted light curves, after cleaning, 
as a function of time. For clarity, the lower panels show zooms on the out-of-eclipse 
modulations (coming from cold spots and detrending procedures), which often
evolve in very short time scales, of single orbital periods. In some cases, the detrending 
algorithm left a discontinuity in the light curve. In the case of A-085 (EPIC~211839462) 
the varying pointing of the telescope resulted in changes of the third light contamination. 
For all these reasons, the majority of light curves were analysed in parts. Their number per 
target and campaign, as well as the total number of data points in a given campaign, are given 
in Table~\ref{tab_lcsum}. In Figure~\ref{fig_lcsum}, different parts are represented
by different colours.

Finally, we would like to note that for A-171 (EPIC~234440875) we only took data from the first 
part of Campaign 11. This campaign was separated into two segments as a result of an error in 
the initial roll-angle used to minimize solar torque on the spacecraft\footnote{More details in 
\url{https://keplerscience.arc.nasa.gov/k2-data-release-notes.html\#k2-campaign-11}}. Data from 
the second part are of worse quality, with significantly larger scatter and short-term systematics, 
which led to larger uncertainties of the resulting parameters. The adopted data set continuously 
covers about five orbital periods, which we find enough for a proper analysis.

\section{Analysis}\label{sec_ana}

\subsection{RV fitting}
The RV solutions were found using the procedure called \vfit \citep{kon10}. We used 
it to fit a double-Keplerian orbit to a set of RV measurements of two components, 
utilizing the Levenberg-Marquardt minimization scheme. The fitted parameters are: orbital 
period $P$, zero-phase moment $T_P$\footnote{Defined in this code as the moment of passing the 
pericentre for eccentric orbits or quadrature for circular.}, systemic velocity $\gamma$, 
velocity semi-amplitudes $K_{1,2}$, eccentricity $e$ and periastron longitude $\omega$.
Depending on the case, we also included the difference between systemic 
velocities of two components, $\gamma_2-\gamma_1$, and difference between zero 
points of different spectrographs. 
Whenever applicable, we simplified our fit by keeping the orbital period on the value given 
by initial fits to complete light curves (see next Section), or by fixing $e$,
$\gamma_2-\gamma_1$, or instrument zero points differences to zero, when any of these 
parameters was found indifferent form 0.0.

Systematic errors that come from fixing a certain parameter in the fit are assessed by a Monte-Carlo 
procedure, which perturbs the value of such parameter within its given error
(e.g. when orbital period is known from light curve analysis).
Other possible systematics (like coming from poor sampling, low number of measurements, 
pulsations, activity etc.) are estimated by a bootstrap analysis. All the uncertainties of orbital 
parameters given in this work already include the systematics.

Moreover, to obtain reliable formal parameter errors of the fit, and the final reduced $\chi^2$ 
to be close to 1, we were modifying the RV measurement errors either by adding a systematic term 
(jitter) in quadrature, or multiplying by a certain factor. Adding the jitter works better for 
active stars, like A-083 or A-231, when the RV scatter is caused by spots, and is compensated 
with the additional term. However, since \vfit weights the measurements on the basis of their own 
errors, which are sensitive to SNR and rotational velocity, we mainly used the 
second option, in which the weights are preserved, and which is more suitable for stars with 
significant flux difference (like A-045).

\subsection{Light curve fitting}\label{sec_lcfit}

The \ktwo light curves were fitted with the version 40 (v40) of the code \jkt 
\citep{sou04a,sou04b}, which is based on the \textsc{ebop} program \citep{pop81}. 
It is designed to work with well-separated binaries, with the ``oblateness''
of components not exceeding 4\%. The highest value in our sample was found to be 1.1\%
(A-060), thus the use of \jkt is justified.

The code fits the period $P$, mid-time of the primary (deeper) minimum $T_0$, 
sum of the fractional radii $r_1 + r_2$ (where $r = R/a$), their ratio $k$, inclination 
$i$, surface brightness ratio $J$, maximum brightness $S$, as well as for $e$ and $\omega$, 
however their final values are actually from \vfit runs, unless stated otherwise. 
Third light contribution $l_3/l_{\rm tot}$, which can be significant in \ktwo data, 
was also initially fitted for, but when it was found indifferent from zero, 
the fit was repeated with fixed $l_3/l_{\rm tot}=0$. It is worth noting that the
detrending algorithms already correct for additional light from nearby sources (e.g. PDC), 
therefore the $l_3$ value does not always have a physical meaning, unless the SAP curve is 
being modelled. In some cases the contamination level can be overestimated, and in the 
fitting process one can even obtain a negative value of $l_3$.

The gravity darkening 
coefficients and bolometric albedos were always kept fixed at the values appropriate for 
stars with convective envelopes \citep[$g = 0.32$, $A = 0.5$; ][]{luc67,ruc69}.
Reflection coefficients were fixed at zero, with the exception of A-171, where they
were treated as free parameters. For the limb darkening (LD) coefficients we used the logarithmic
law of \citep{kli70}. Initial values were found on the basis of $\log(g)$ found from first 
\vfit and \jkt runs, and temperatures from approximate isochrones (of solar metallicity).
In the final runs, the LD coefficients were fine-tuned during a fit, or at least perturbed 
during the error estimation stage. In some cases, setting the LD free led to physically
impossible values, which might be explained for example by spots affecting the shape of an
eclipse.

\begin{figure*}
    \centering
    \includegraphics[height=5cm]{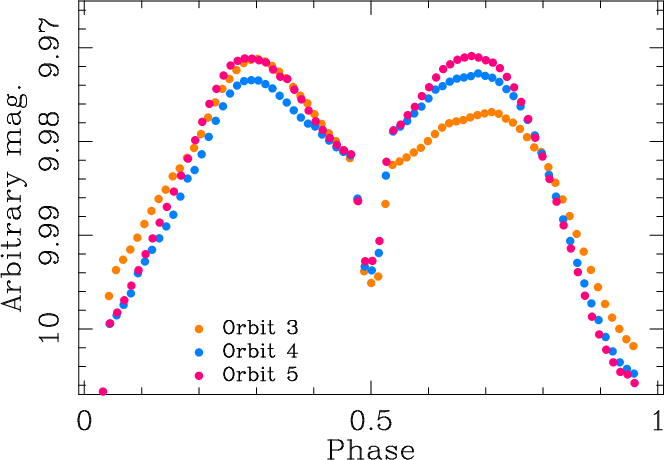}
    \includegraphics[height=5cm]{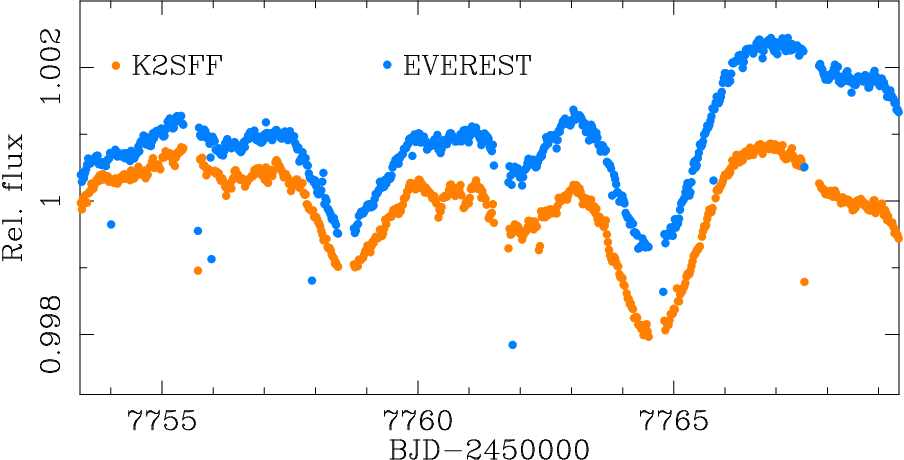}
    \caption{Left: A fragment of the C13 light curve of A-045, limited to three orbital periods (4.96~d), phase-folded, and zoomed to better show the evolution of the out-of-eclipse modulation. Each orbit is shown with a different colour. The change in the spot-originated modulation between consecutive orbits is clearly seen.
    Right: Comparison of two HLSP products (before the removal of outliers) for C12 LC data of A-231, also zoomed to better show the out-of-eclipse modulation. The use of different algorithms resulted in a different shape of the light curve, which may lead to different results when fitting for spots (e.g. the drop in {\sc k2sff} flux could be interpreted as a drop in spot's temperature, increase of its size, or its migration closer to the equator).}
    \label{fig_spotrend}
\end{figure*}

Undoubtedly, the biggest challenge in the light curve fitting in this work was the treatment 
of spots and their influence on the resulting binary parameters. Several systems in our sample show 
stellar spots that are not only prominent, but also change rapidly, with the time scale of
single orbital periods. This evolution is clearly seen 
in high-precision space-borne data \citep[see also: ][]{gil17}. 
This means that the observed pattern of spots varies between two consecutive orbits, 
and is not stable during one orbital cycle (Figure~\ref{fig_spotrend}, left). In other words, the binary 
appears different at the beginning and end of a single period. The scheme of light curve
fitting typically requires (and assumes) that the studied system is stable during at least one
orbital cycle, but one can see that it is not always the case. 

One should also keep in mind, that the model of spots used in various light curve fitting codes 
(if incorporated) is simplified. A single cold spot is usually represented by a circular area of lower
surface brightness, parametrised by its size, position, and contrast with respect to the ``clear'' 
photosphere. Multiple spots are often used in the analysis. The problem of modelling a spot 
distribution on a two-dimensional surface into a one-dimensional light curve is, however, ill-posed, 
i.e. there is more than one model that reproduces the observed light curve 
\citep[e.g.][]{win10,ioa16,bas20}, and different spot distributions (e.g. changing spot's 
location from primary to secondary) leads to variations in resulting 
stellar parameters at the level of several per-cent \citep[e.g.][]{win10,hel11}. Even under this
simplified spot model, the basic parameters are degenerated, i.e. change in spot's size can have the same
effects on the light curve as the change in its temperature and/or latitude \citep{lan04,ioa16}.
In some cases, the use of multi-band photometry could help, but it would have to be taken simultaneously, 
and the current space photometric missions do not have such capabilities. Additionally,
the observational data suggest that a significant population of short-period DEBs, which should be 
tidally locked, shows differences in rotation velocity with latitude \citep{lur17}.

To properly model the rapidly-varying spots, as seen in our cases, one needs a code that allows for
time evolution of all the parameters, and most likely non-linear. The later versions of the
Wilson-Devinney code \citep[WD;][]{wil71,wil12} allow for migration of spots in longitude and 
their growth and decay (spot aging) in size. Still, the modelling process remains complicated, 
requires a lot of attention and time (as different distributions of spots need to be tested),
does not allow for latitudal differential rotation, and possibility of failure (no convergence 
if starting from inaccurate values, or finding the local minimum) is high.

On top of that, the shape of the out-of-eclipse modulation can also be altered by the detrending
scheme that was used in the data preparation (Figure~\ref{fig_spotrend}, right). For example, a gradual
drop in brightness, that was in fact introduced by the algorithm, could be mistaken with the
increase of the spot's size, or decrease of its surface brightness. On the other hand, artificial
``flattening'' of the light curve leads to false stability of the parameters, and causes the loss 
of information about the true intrinsic variation. Inaccurate assumptions about distribution
of spots and, simultaneously, instrumental effects may lead to, for example, incorrect
depths of eclipses, which would affect temperature ratios, fractional radii and inclination.

In case of space-borne photometry of heavily spotted DEBs, one deals with a combination
of {\it a priori} unknown effects, both intrinsic and artificial, and can not be sure
how the resulting parameters of spots (and stars) were affected. Even if the migration 
and aging options are included (like in the WD), one still has to deal with several 
issues: (i) the size-temperature(-latitude) correlation; (ii) unknown number and location
of spots; (iii) possible differential rotation; (iv) imperfect detrending; (v) time required.
In such case it is very difficult to assess reliable uncertainties of the results. 
To our knowledge, there are no light curve fitting codes that allow to undertake 
this kind of analysis in a satisfactory and efficient way. A promising way to overcome
this may be modelling the activity signals with Gaussian Processes (GP), implemented for
example in \citet{gil14}, \citet{gil17} or \citet{smi21}. The advantage is that almost any
out-of-eclipse modulation can be modelled simultaneously with the eclipsing binary, 
and uncertainties of the GP regression can be propagated to the stellar parameters. The dynamical 
character of the spots is therefore accounted for without modelling a spot {\it per se}, 
and at least partially, included in the error budget. It is, however, not clear, if 
the ambiguity of the process (i.e. applying the GP to the total brightness of the
system, instead of just one component that could have spots) is also properly 
reflected in the final errors of stellar parameters.

For these reasons we decided to apply a different approach, with similar 
foundations as the one by \citet{gil17}. We do not try to model the spots as physical 
objects, but instead we  embrace the fact that we have our results affected by 
systematics, and focus on the proper and reliable determination thereof in a time-efficient 
way. We base our approach on the fact that the spots evolve quickly, and the light curve 
looks different after single orbital periods. 

As mentioned before, \ktwo data were often divided into shorter parts, to easier deal with the
evolution of spots, changes in the level of $l_3$, or other instrumental effects. Dividing
data into subsets has been used or suggested as a robust way to obtain parameter errors by e.g.
\citet{max18a,max20} or \citet{hel19b}. The \jkt v40
is capable of fitting a number of polynomials (up to the 5th order) and sine functions to 
compensate for variations in some of the parameters (e.g. $l_1, l_2, l_{\rm tot})$. 
We used these features to model the dynamical out-of-eclipse variations coming from spots 
and long-term instrumental trends, as well as putative pulsations.
The sines were found through an iterative process. After each \jkt run, a Lomb-Scargle periodogram 
of the residuals was made, and the most prominent frequency was identified. The
next fit included a sine function with the recently found frequency. If the new fit was 
better than the previous (in terms of $rms$), the frequency was kept in the model, otherwise the 
second most prominent frequency was taken. These steps were repeated until no further advance in 
the quality of the overall model was noted. The number of sines and polynomials used varied 
between targets and parts. The exact values of the polynomial coefficients, and sine periods and
amplitudes are given in Table~\ref{tab_pierdy} of the Appendix.

Parameter errors were estimated in two steps. First, for each single part we either 
used the the Monte-Carlo (MC) method or the residual-shift (RS) approach \citep{sou11},
both available in \jkt (tasks 8 and 9, respectively). The MC was preferred for light curves 
highly variable in time (like A-045 or A-085), while the RS was applied for more stable cases
(A-060 or A-171) where short-term instrumental effects contribute more. These errors were adopted 
as final ones when single cadence/part was analysed (i.e. A-060, A-163, and A-171), or used 
for weighting, in case of multiple cadences/parts. In such cases, the adopted error was created
by adding in quadrature the formal error of the weighted average and the $rms$ of individual 
values \citep{hel15}. In general, when the light curve shape changes rapidly the $rms$ 
term dominates over the average, and {\it vice versa} for stable cases.
In this way we take into account the influence of changes in the shape
of the light curve (of any origin) into the uncertainties of a given parameter.

The only exception to the scheme above were the errors of $P$ and $T_0$, which were found with
the MC option (task 8) used on complete sets of data.

\subsection{Spectra disentangling}
We used our spectroscopic data and run a disentangling procedure in order to obtain 
separate spectra of the components, suitable for further direct determination of 
effective temperatures and metallicities. We applied the version 3 of the code {\sc fdbinary}
\citep[{\sc fd3};][]{ili04}\footnote{\url{http://sail.zpf.fer.hr/fdbinary/}}, 
which performs separation of spectra in the Fourier space. This code has been chosen because
it is capable of disentangling three components, which was needed in the case of A-045.
Because \fdb uses component fractional intensities as input, we performed 
the separation in a relatively small wavelength range, where the flux ratio is roughly 
constant. We chose the range 5000--5500~\AA, rich in stellar spectral features, not affected by 
tellurics, and in which the SNR of individual spectra was relatively good. At the end, the SNR for 
each component spectrum was evaluated on the basis of the SNRs of individual observations, 
and intensity ratio of components (from \todcor).

In the \fdb runs we did not 
combine spectra from different instruments. In particular, HIDES data were used for 
A-045 and A-083 = RU~Cnc, CHIRON-fiber for A-060 and A-085, and FEROS for A-163, A171, and 
A-231. The remaining A-111 = FM~Leo was not treated with this scheme, as its disentangled spectra
(from CHIRON-slicer) were obtained by \citet{syb18} with the tomographic approach described
in \citet{kon10}.

\subsection{Spectroscopic analysis}
To obtain individual effective temperatures, and systemic metallicities from the decomposed
spectra, we used the v2020.10.01 version of the freely distributed code \ispec \citep{bla14}.
Flux errors were introduced on the basis of the previously calculated SNR. 
In this way we made sure the resulting uncertainties are trustworthy, which was verified 
by the reduced $\chi^2$, given in the output.

To find the atmospheric parameters we used the spectral synthesis approach, utilising 
the code {\sc spectrum} \citep{gra94}, the MARCS grid of model atmospheres \citep{gus08}, 
and solar abundances from \citet{gre07}. \ispec synthesizes spectra only in certain,
user-defined ranges, called ``segments''. We followed the default approach, where these 
segments are defined as regions $\pm$2.5~\AA\,around a certain line. We decided to synthesize
spectra around a set of lines carefully selected in such way, that various spectral 
fitting codes reproduce consistent parameters from a reference solar spectrum \citep{bla16}.

We run fits with the following parameters
set free: effective temperature $T_{\rm eff}$, metallicity [M/H], alpha enhancement 
[$\alpha$/Fe], and microturbulence velocity $v_{\rm mic}$. The resolution $R$ was
always fixed to a value appropriate for a given instrument, and gravity $\log(g)$ to 
the value corresponding to absolute values of mass and radius (see next Section), 
which is more precise than $\log(g)$ found from spectroscopy. 
The rotational velocity $v\sin(i)$ was also fixed and set to values expected from 
the synchronous rotation, which is expected for short-period circular or nearly circular
orbits, as typically in our sample. In two cases, however -- A-060
and A-163 -- we set $v\sin(i)$ free, for reasons that will be explained later in the text. 
The macroturbulence velocity $v_{\rm mac}$, which degenerates with rotation, was at all 
times calculated on-the-fly by \ispec from an empirical relation.

As final values of systemic [M/H] and [$\alpha$/Fe] we adopted averages of values obtained
from each component. As their conservative uncertainties we added in quadrature the average 
formal parameter errors from \ispec and standard deviation of the individual results. 
It is worth noting, that the values of [$\alpha$/Fe] were all formally indifferent
from zero. In one case -- A-171, the hottest system in our sample -- the [$\alpha$/Fe] was
found to be -1.21$\pm$1.03~dex, which we find suspiciously low. Therefore we repeated the 
analysis with [$\alpha$/Fe] fixed at 0.

In case of A-043 and A-085 we could also analyse spectra other than of the two eclipsing 
components. For the former, we actually run \ispec on the primary and tertiary components 
only, as the secondary contributed less than 1\% to the total flux, and the SNR of the 
disentangled spectrum was only $\sim$2. For the latter, we also obtained one CHIRON-fiber
observation during the total primary eclipse (when only the secondary's light was recorded).
Additionally, early CORALIE and HDS observations of the visual companion were also used.
In \ispec runs for both tertiaries the $\log(g)$ parameter was set free. For both A-045 
and A-085, the final values of systemic [M/H] and [$\alpha$/Fe] were derived from all
available spectra.

\subsection{Absolute stellar parameters}
The absolute values of stellar parameters were calculated with the 
{\sc jktabsdim}\footnote{\url{http://www.astro.keele.ac.uk/jkt/codes/jktabsdim.html}}
procedure, which is available with {\sc jktebop}. This code combines the output of spectroscopic 
and light curve solutions to derive a set of stellar absolute dimensions, related quantities, 
and distance, if effective temperatures are given. For this purpose the code uses the 
apparent, total magnitudes of a given binary in any of the $U,B,V,R,I,J,H,K$ bands. 
It compares the observed (total) magnitudes with absolute ones, calculated using a number of 
bolometric corrections \citep{bes98,cod76,flo96,gir02}, and surface brightness-$T_\mathrm{eff}$ 
relations from \citet{ker04}. 
Flux ratios may also be used to further constrain individual 
absolute magnitudes of each component.

Apart from stellar, photometric, and orbital parameters, \jktabs also 
calculates the rotation velocities predicted for the case of synchronisation 
of rotation with orbital period $v_{\rm syn}$, the time scale of such synchronisation
and the time scale of circularisation of the orbit.

\subsection{Isochrones}\label{sec_iso}

The age of each system, and evolutionary status of each star were estimated 
with a grid of isochrones generated using a dedicated web
interface,\footnote{\url{http://waps.cfa.harvard.edu/MIST/}} based on the 
Modules for Experiments in Stellar Astrophysics \citep[MESA;][]{pax11,pax13,pax15,pax18}, 
and developed as part of the MESA Isochrones and Stellar Tracks project 
\citep[MIST v1.2;][]{cho16,dot16}. For the majority of the star's evolution 
in a DEB, the only interaction between the components is through gravity, 
which becomes important for small separations, leading to tidal interactions and 
consecutive phenomena (synchronisation, alignment, faster rotation, etc).
The MIST models include rotation by default, but do not incorporate the 
{\sc MESAbinary} module, that is capable of evolving structure of two stars \citep{pax15}.
However, for the main sequence, early RGB phases and masses well below 2~M$_\odot$, 
the differences in stellar parameters are smaller than our measurement errors, 
and do not affect the resulting ages significantly. The activity may be, 
however, more relevant, at least for some of our objects (like A-045), as it
can strongly affect the radii and effective temperatures of low-mass and sub-giant
stars. The most recent isochrones for active stars \citep{som20} can more reliably
reproduce those properties. We therefore conclude that in at least one case -- A-045
-- our age determination might be affected by the activity level.

The grid of isochrones was generated for iron abundance 
[Fe/H]\footnote{It is reasonable to assume that without significant deviations from solar
amounts of $\alpha$-elements, the iron abundance [Fe/H] sufficiently approximates the 
amount of elements metals [M/H].} values from -4.0 to 0.50~dex 
with 0.05~dex steps, as well as for ages 10$^{8.6}$ to 10$^{10.2}$~Gyr,
in logarithmic scale, every $\log(\tau)$=0.01.

On each isochrone we were looking for a pair of points that simultaneously best reproduce the 
observed masses $M_{1,2}$, radii $R_{1,2}$, and effective temperatures $T_{\rm eff1,2}$ 
of two components, as well as their flux ratio $l_2/l_1$ in the \kep band, metallicity [Fe/H],
and GEDR3 distance. The reddening-free distance $d_0$ was estimated simultaneously with the 
reddening $E(B-V)$, using the available observed total magnitudes in different filters, and 
the predicted total brightness of the system in the same filters, for a given pair of points
(= stellar masses). 
Distances in each available band $d_\lambda$ were calculated from $T_{\rm_eff}$--surface
brightness relations from \citet{ker04}. Then, they were transformed to distance moduli
$(m-M)_\lambda$ with the standard relation $(m-M) = 5 \log(d) + 5$. Due to the interstellar
extinction and reddening, individual values of $(m-M)_\lambda$ were obviously not in agreement.
To obtain the extinction-free modulus $(m-M)_0$ we fitted a straight line on the $A_\lambda$
vs. $(m-M)_\lambda$ plane, where the $A_\lambda$ are extinction coefficients in each band. 
We followed the extinction law of \citet{car89}
$A_U:A_B:A_V:A_R:A_I:A_J:A_H:A_K = 4.855:4.064:3.1:2.545:1.801:0.88:0.558:0.36$,
which assumes $R_V=3.1$.
The slope of the fitted line in this approach is the reddening $E(B-V)$, while the
intercept is the extinction-free modulus $(m-M)_0$, which can be translated into the distance
$d_0$. Additionally, when spectroscopic results for the tertiaries were available (i.e. their
$T_{\rm eff}$ and $\log(g)$), we also verified if they are well reproduced by the 
isochrone. This allowed us to constrain their other properties, like masses, radii, and 
evolutionary status.

It is worth noting that, like any other measurement, the [M/H] values from \ispec are
uncertain, thus looking for the best-fitting isochrone, while keeping [M/H] fixed, is not
the most optimal approach. The values of [Fe/H] (assumed equal to [M/H], since 
[$\alpha$/Fe]=0) given as the result of isochrone fitting, may not be the same as 
those found in {\sc ispec}. Hereafter, [M/H] refers to the \ispec results, while 
[Fe/H] to isochrones.

\section{Results}\label{sec_res}

\begin{figure*}
    \centering
	\includegraphics[width=0.32\textwidth]{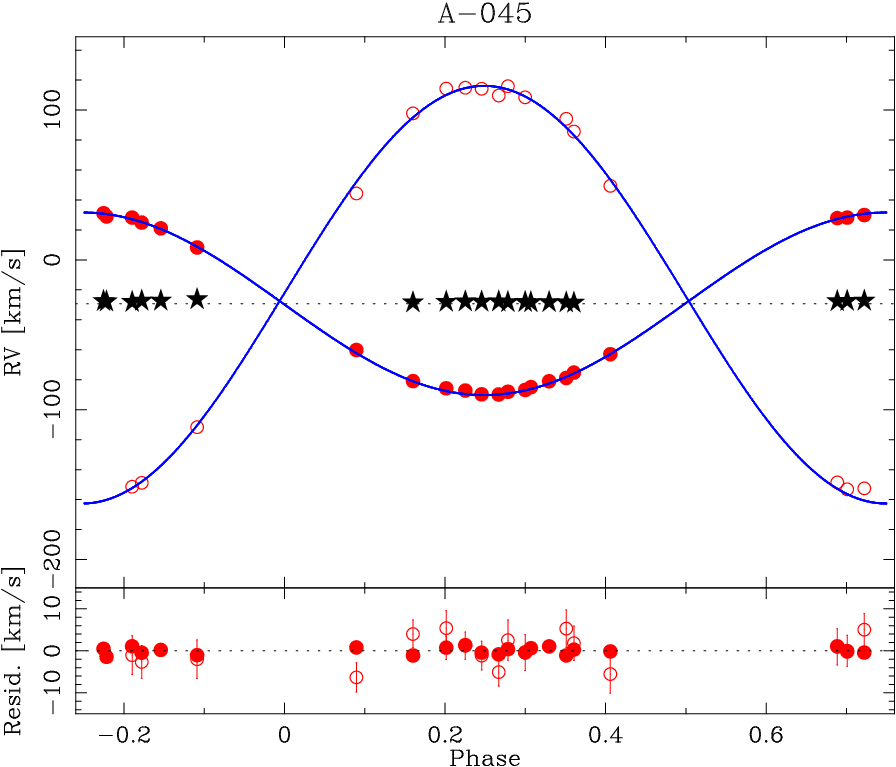}
	\includegraphics[width=0.32\textwidth]{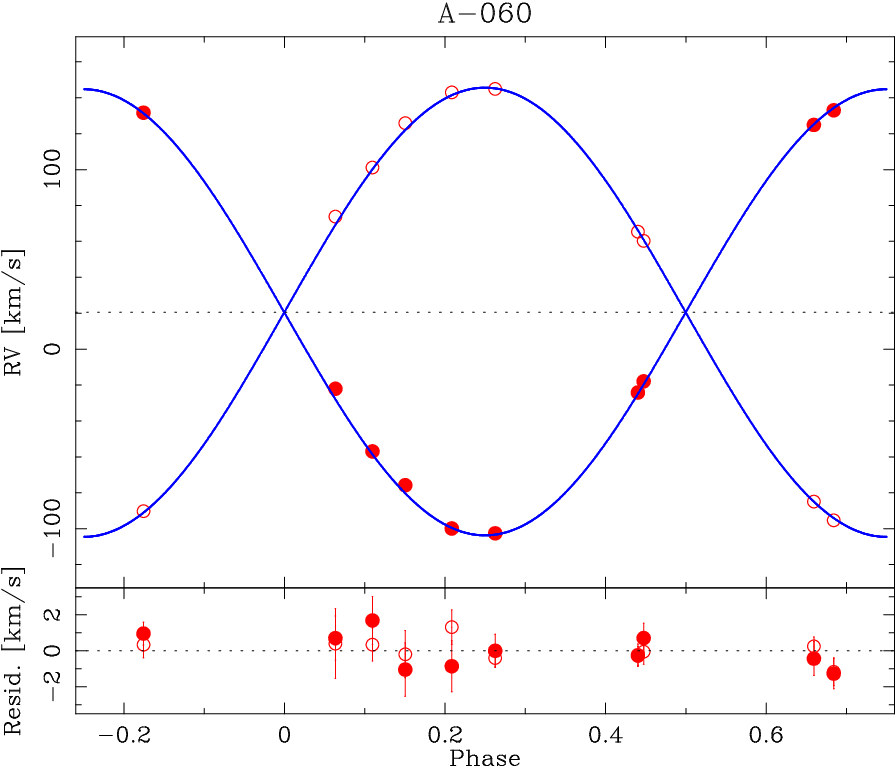}
	\includegraphics[width=0.32\textwidth]{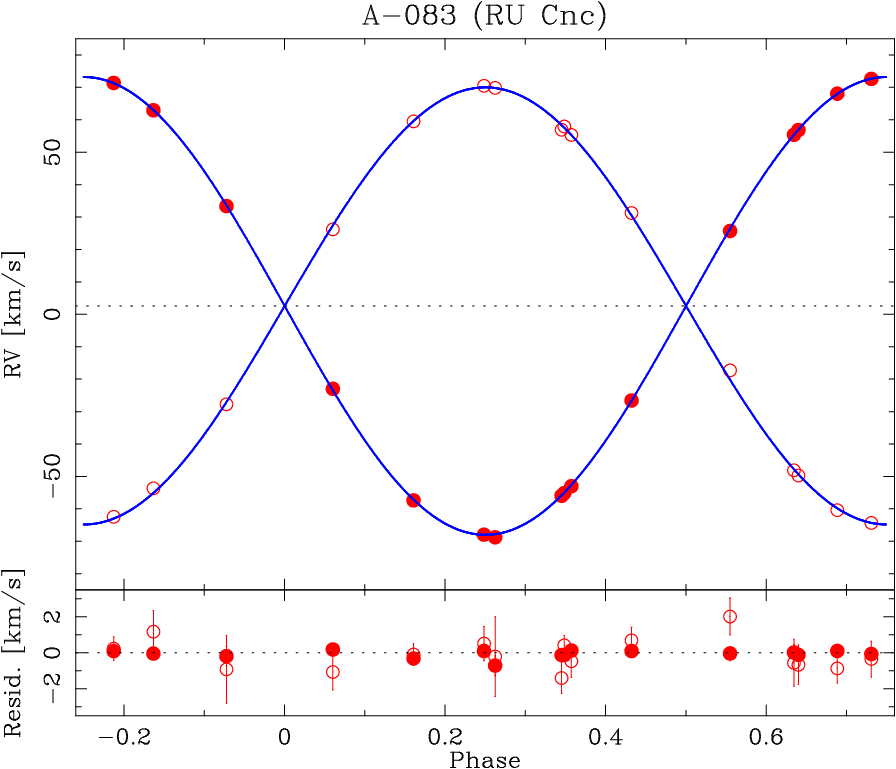}
	\includegraphics[width=0.32\textwidth]{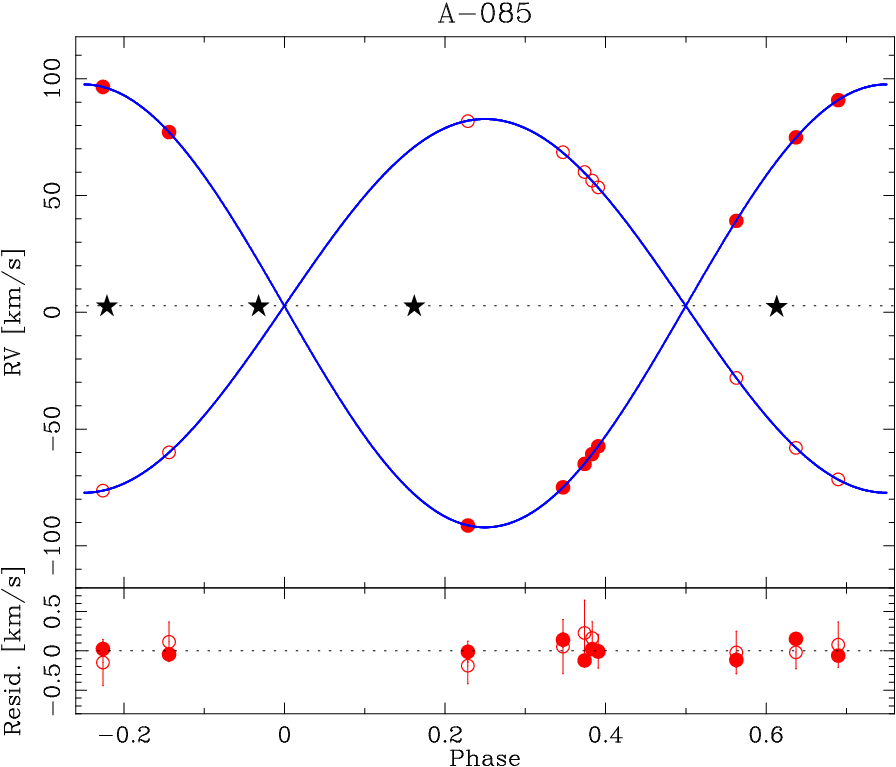}
	\includegraphics[width=0.32\textwidth]{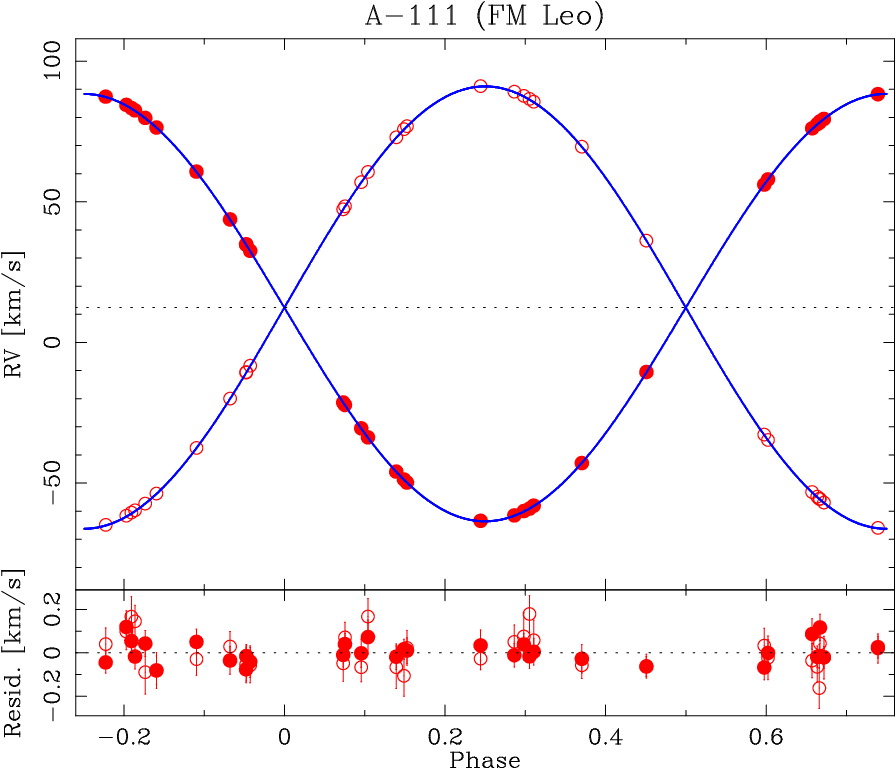}
	\includegraphics[width=0.32\textwidth]{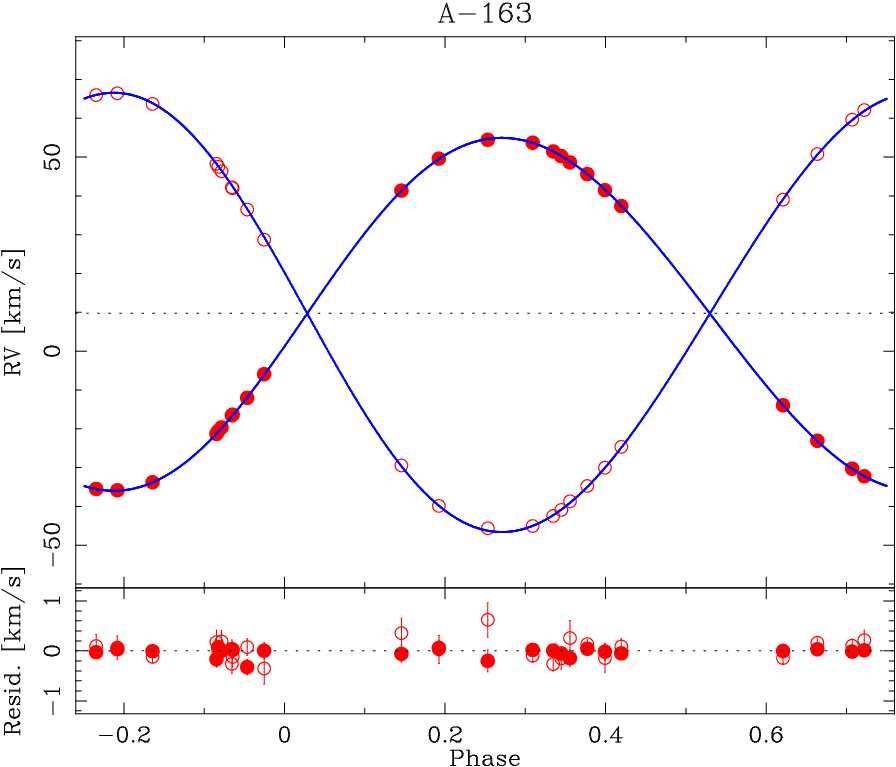}
	\includegraphics[width=0.32\textwidth]{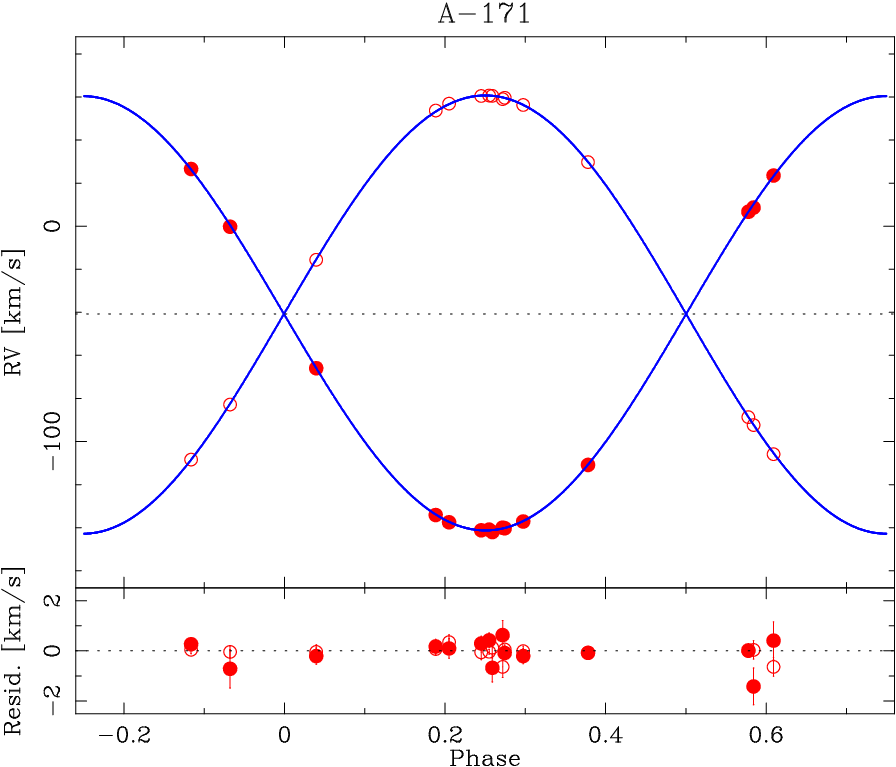}
	\includegraphics[width=0.32\textwidth]{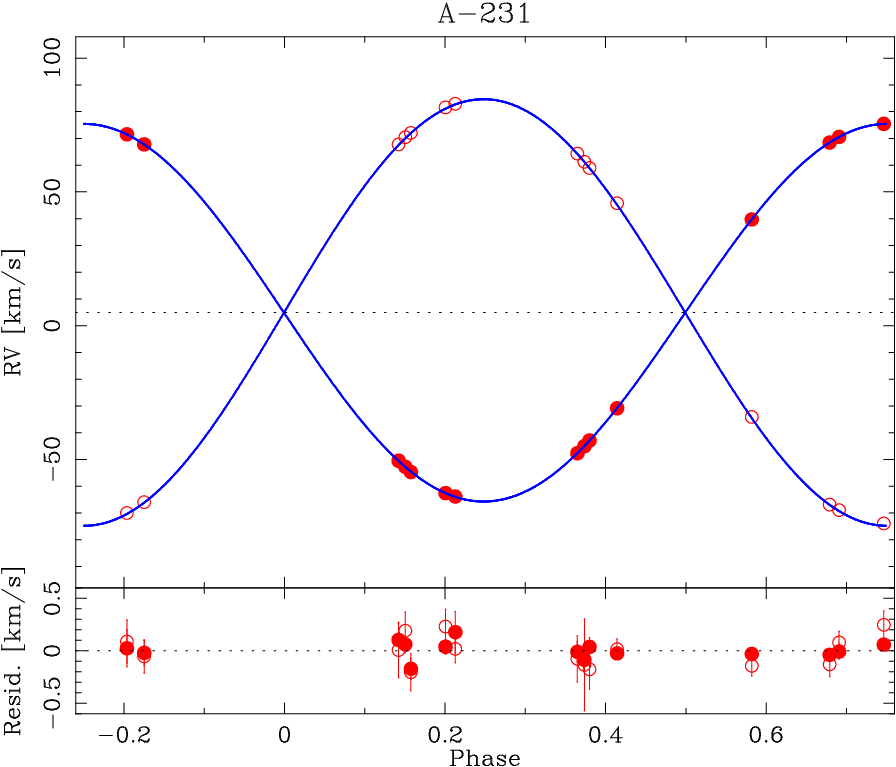}
    \caption{RV measurements (red points) and solutions (blue lines) for the studied systems,
    phase-folded with the orbital period, with phase 0 set to the moment of the
    primary eclipse. Filled points are for the primaries, and open for secondaries. Black dashed 
    horizontal lines mark systemic velocities $\gamma$. Lower panels show residuals of fits. 
    Black stars on the plots for A-045 and A-085 represent RVs of the companions
    to the DEB.}
    \label{fig_rv}
\end{figure*}

Our final models are presented in Figures~\ref{fig_rv} (RVs) and \ref{fig_lc1} (light curves). 
The best-fitting isochrones are shown together with our 
measurements on the $M-R$ and $M-T_{\rm eff}$ planes in Figure~\ref{fig_iso}. 
The results, including orbital, physical, and atmospheric stellar parameters, as
well as age and $E(B-V)$, are listed in Table~\ref{tab_par}. 

Below we briefly discuss our results for each system separately.

\subsection{A-045}

This old, ($\sim$9.3 Gyr) but highly active binary is composed of a nearly-solar-mass 
primary and a low-mass secondary. This is in general an interesting configuration 
-- such pairs, composed of two vastly different stars, allow for more stringent tests 
of stellar evolution models than pairs of nearly identical components. For a given
metallicity, their ages are tightly constrained by their masses and radii.

\begin{landscape}
\begin{table}
\centering
\caption{Orbital, physical, and atmospheric parameters of the studied systems. Values that were fixed and automatically calculated with empirical calibrations, are denoted by ``fix'' and ``emp'', respectively.}\label{tab_par}
\begin{tabular}{lcccccccc}
\hline \hline
ASAS ID                 & 045021+2300.4 & 060505+2032.1 & 083730+2333.7 & 085002+1752.5 & 111245+0020.9 & 163903-2847.2 & 171750-1915.3 & 231922-0852.2 \\
EPIC                    & 247605441		& 202073040     & 212173112		& 211839462     & 201488365     & 202674012		& 234440875     & 246024234     \\
Running name            & A-045         & A-060         & A-083 (RU Cnc)& A-085         & A-111 (FM Leo)& A-163         & A-171         & A-231         \\
\hline
$P$ (d)                 & 1.653362(7)   & 2.1212341(35) & 10.1729311(55)& 5.22569411(9) & 6.728606(2)   & 23.309595(35) & 3.136930(13)  & 6.062036(4)   \\
$T_0$ (JD-2450000)$^a$  & 7020.57260(11)& 6768.11068(4) & 7747.34574(14)& 7143.61747(14)& 6812.22376(1) &6909.36019(12) & 7660.70357(3) &7743.442277(32)\\
$T_P$ (JD-2450000)$^b$  & 7021.8138(21) & 6678.4861(14) & 7053.0440(35) & 6677.2237(06) & 7429.5730(3)  & 6081.11(10)   & 6427.1050(4)  & 6106.92(37)   \\
$K_1$ (k\ms)            &   60.33(25)   & 123.7(1.0)    &  70.62(9)     &  94.86(5)     &  76.033(19)   &  45.44(3)     & 100.66(12)    &  70.59(5)     \\
$K_2$ (k\ms)            &  137.7(1.0)   & 125.1(6)      &  67.2(3)      &  79.99(8)     &  78.646(22)   &  56.56(7)     & 101.55(8)     &  79.71(11)    \\
$\gamma_1$ (k\ms)       &  -29.0(2)     &  20.7(5)      &   2.55(9)     &   2.73(4)     &  12.527(51)   &   9.70(5)     & -40.76(6)     &   4.90(6)     \\
$\gamma_2-\gamma_1$ (k\ms)& 6.8(1.0)    &   0.0(fix)    &   0.2(3)      &   0.0(fix)    &   0.036(33)   &   0.0(fix)    &   0.0(fix)    &   0.21(22)    \\
$q$                     &   0.4383(37)  &   0.989(9)    &   1.050(4)    &   1.1859(14)  &   0.9668(4)   &   0.8033(12)  &   0.9912(13)  &   0.8855(14)  \\
$M_1\sin^3(i)$ (M$_\odot$)& 0.924(17)   &   1.702(22)   &   1.347(11)   &   1.324(3)    &   1.3119(8)   &   1.420(4)    &   1.3495(26)  &   1.131(3)    \\
$M_2\sin^3(i)$ (M$_\odot$)& 0.405(5)    &   1.683(27)   &   1.415(7)    &   1.570(2)    &   1.2683(7)   &   1.140(2)    &   1.3376(32)  &   1.002(2)    \\
$a\sin(i)$ (R$_\odot$)  &   6.472(34)   &  10.43(5)     &  27.73(6)     &  18.066(10)   &  20.573(4)    &  46.99(4)     &  12.541(8)    &  18.015(14)   \\
$e$                     &   0.0(fix)    &   0.0(fix)    &   0.0(fix)    &   0.0(fix)    &   0.0(fix)    &   0.026(2)    &   0.0030(15)  &   0.0049(33)  \\
$\omega$($^\circ$)      &   ---         &   ---         &   ---         &   ---         &   ---         & 259(1)        & 272(2)        & 103(12)       \\
$r_1$                   &   0.1629(34)  &   0.1957(4)   &  0.0789(15)   &   0.0869(13)  &   0.07900(43) &   0.04588(12) &   0.1334(5)   &   0.0909(5)    \\
$r_2$                   &   0.0628(14)  &   0.1919(5)   &  0.1781(26)   &   0.1783(24)  &   0.07293(58) &   0.02603(8)  &   0.1312(5)   &   0.0615(13)   \\
$i$ ($^\circ$)          &  85.26(36)    &  85.81(2)     &  89(1)        &  88.1(7)      &  87.939(28)   &  88.668(14)   &  86.486(14)   &  87.01(19)    \\
$J$                     &   0.08(2)     &   0.996(4)    &   0.205(12)   &   0.624(21)   &   0.996(7)    &   0.965(12)   &   0.981(8)    &   0.892(33)   \\
$l_2/l_1$               &   0.011(3)    &   0.954(9)    &   1.10(8)     &   2.58(15)    &   0.843(22)   &   0.3112(27)  &   0.975(11)   &   0.405(17)   \\
${l_3/l_{\rm tot}}^c$   &   0.0(fix)    &   0.016(3)    &   0.0(fix)    &   variable    &   0.007(4)    &   0.013(7)    &   0.051(4)    &   0.031(19)   \\
$rms_{\rm RV1}$ (k\ms)  &   0.83        &   0.97        &   0.23        &   0.09        &   0.051       &   0.10        &   0.54        &   0.082       \\
$rms_{\rm RV2}$ (k\ms)  &   3.71        &   0.65        &   0.90        &   0.13        &   0.084       &   0.22        &   0.26        &   0.147       \\
$rms_{\rm LC}$ (mmag)   &   0.97        &   0.84        &   1.1         &   0.28        &   0.16        &   0.19        &   0.33        &   0.20        \\
$M_1$ (M$_\odot$)       &   0.934(17)   &   1.716(22)   &   1.347(11)   &   1.326(3)    &   1.3144(8)   &   1.421(4)    &   1.3571(26)  &   1.136(3)    \\
$M_2$ (M$_\odot$)       &   0.409(5)    &   1.697(28)   &   1.415(7)    &   1.573(3)    &   1.2707(7)   &   1.141(2)    &   1.3452(32)  &   1.006(2)    \\
$a$   (R$_\odot$)       &   6.494(35)   &  10.46(5)     &  27.73(6)     &  18.076(12)   &  20.591(4)    &  47.00(4)     &  12.564(9)    &  18.040(15)   \\
$R_1$ (R$_\odot$)       &   1.058(23)   &   2.048(10)   &   2.188(42)   &   1.571(25)   &   1.627(9)    &   2.157(6)    &   1.676(7)    &   1.640(9)    \\
$R_2$ (R$_\odot$)       &   0.408(9)    &   2.008(11)   &   4.939(73)   &   3.222(42)   &   1.498(12)   &   1.224(4)    &   1.648(6)    &   1.110(23)   \\
$\log(g_1)$             &   4.360(18)   &   4.050(3)    &   3.888(17)   &   4.168(14)   &   4.134(5)    &   3.923(2)    &   4.122(3)    &   4.064(5)    \\
$\log(g_2)$             &   4.829(19)   &   4.062(4)    &   3.202(13)   &   3.619(11)   &   4.191(7)    &   4.320(3)    &   4.133(3)    &   4.350(18)   \\
$v_{\rm rot,1}$	(k\ms)$^d$&32.4(7)s     &  39.7(6.6)i   &  10.9(2)s     &  15.21(24)s   &  12.23(7)s    &   14.1(7)i    &  27.03(10)s   &  13.69(7)s    \\
$v_{\rm rot,2}$	(k\ms)$^d$&12.5(3)s     &  35.3(6.2)i   &  24.6(4)s     &  31.18(42)s   &  11.26(9)s    &   5.7(5.8)i   &  26.57(10)s   &   9.26(19)s   \\
$T_{\rm eff,1}$ (K)     &   5668(71)    &   7405(432)   &   6569(84)    &   6554(207)   &   6371(115)   &   6630(83)    &   6530(130)   &   6263(41)    \\
$T_{\rm eff,2}$ (K)     & 3590(100)$^e$ &   7407(382)   &   4761(147)   &   5730(250)   &   6353(116)   &   6586(380)   &   6501(142)   &   6240(99)    \\
$[M/H]^f$               &   -0.26(26)   &   -0.12(21)   &   -0.26(34)   &   -0.01(27)   &   -0.13(9)    &   -0.19(16)   &    0.04(12)   &   -0.39(7)    \\
$[\alpha/Fe]$           &    0.03(63)   &    0.00(fix)  &   -0.04(37)   &   -0.02(10)   &    0.10(23)   &    0.03(15)   &   -0.03(15)   &    0.10(7)    \\
$v_{\rm mic,1}$ (k\ms)  &    2.42(22)   &    2.55(80)   &   10.07(emp)  &    1.45(31)   &    1.60(19)   &    2.16(14)   &    2.27(31)   &    1.91(9)    \\
$v_{\rm mic,2}$ (k\ms)  &   ---         &    2.44(76)   &    2.64(40)   &    1.30(39)   &    1.68(20)   &    2.24(75)   &    2.12(31)   &    0.94(27)   \\
$E(B-V)^g$ (mag)        &   0.132(21)   &   0.163(62)   &   0.201(24)   &    0.004(3)   &   0.054(22)   &   0.139(13)   &   0.217(24)   &   0.037(11)    \\
$\tau$ (Gyr)$^e$            &   9.3(1.5)    &   1.00(15)     &   3.02(40)    &   2.40(28)     &   2.72(31)    &   2.40(30)   &   2.29(34)    &   4.68(57)    \\
\hline 
\end{tabular}\\
$^a$ Mid-time of the primary (deeper) eclipse. $^b$ Time of pericentre or quadrature. 
$^c$ Value obtained in the fit, not bearing a physical meaning (see Sect. \ref{sec_lcfit}).\\
$^d$ Marked with ``s'' for (pseudo-)synchronous rotation velocities, as given by {\sc jktabsdim}, and with ``i'' for projected $v\sin(i)$ obtained with {\sc ispec}.\\
$^e$ Estimated from isochrone fitting. $^f$ From {\sc ispec}. $^g$ Value that reproduces the GEDR3 distance.\\
\end{table}

\end{landscape}

\begin{figure*}
    \centering
	\includegraphics[width=0.32\textwidth]{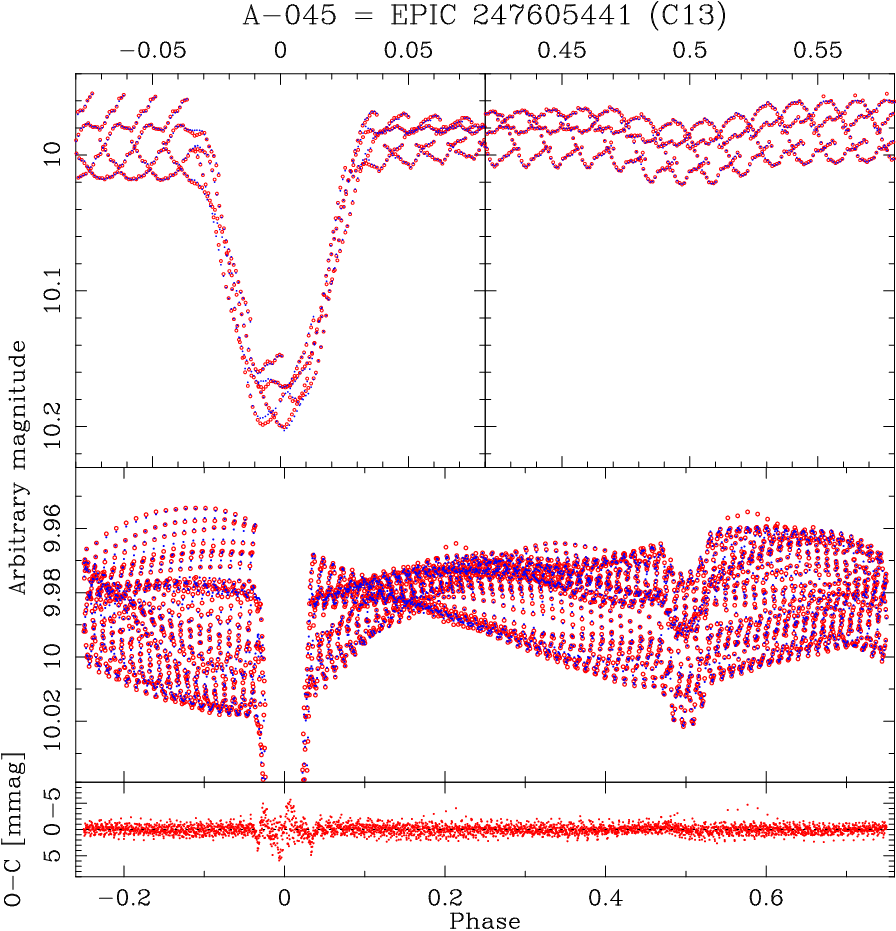}
	\includegraphics[width=0.32\textwidth]{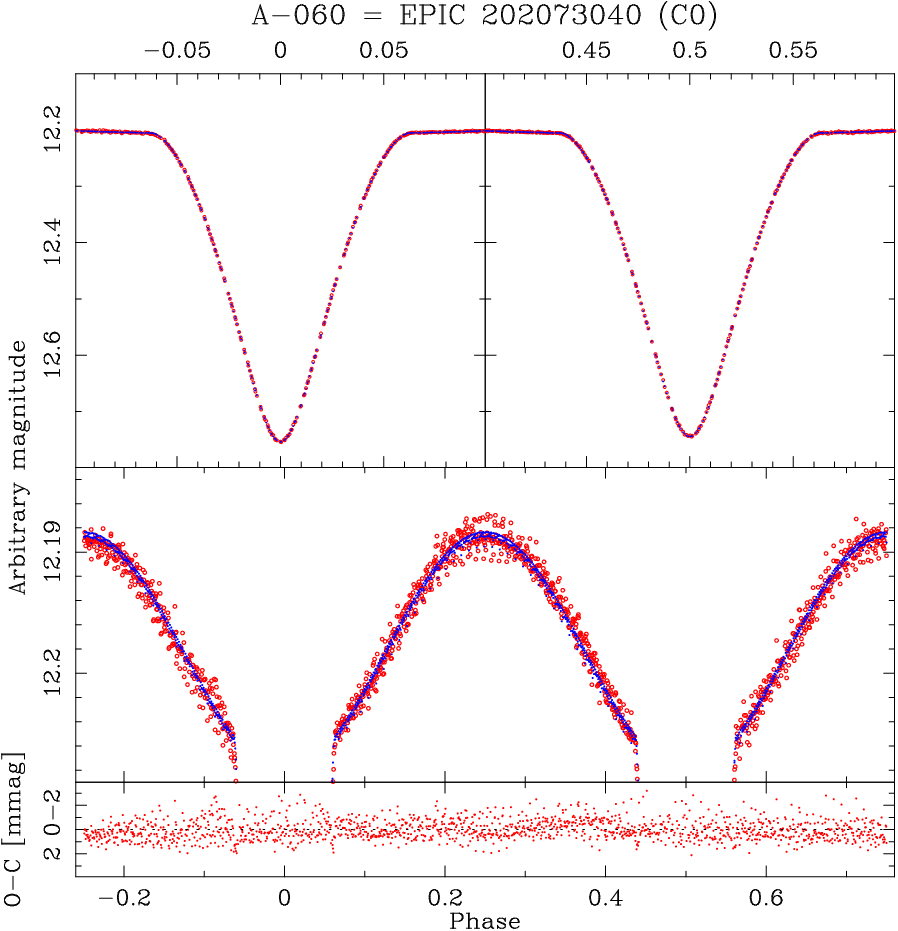}
	\includegraphics[width=0.32\textwidth]{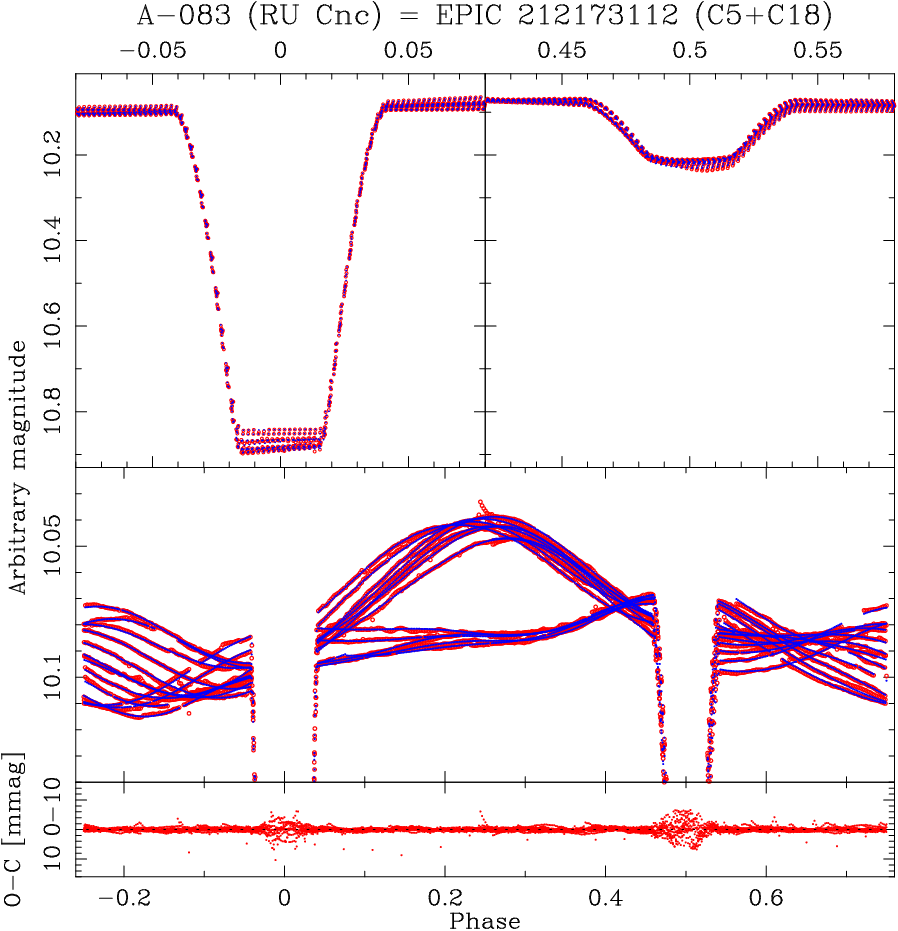}
	\includegraphics[width=0.32\textwidth]{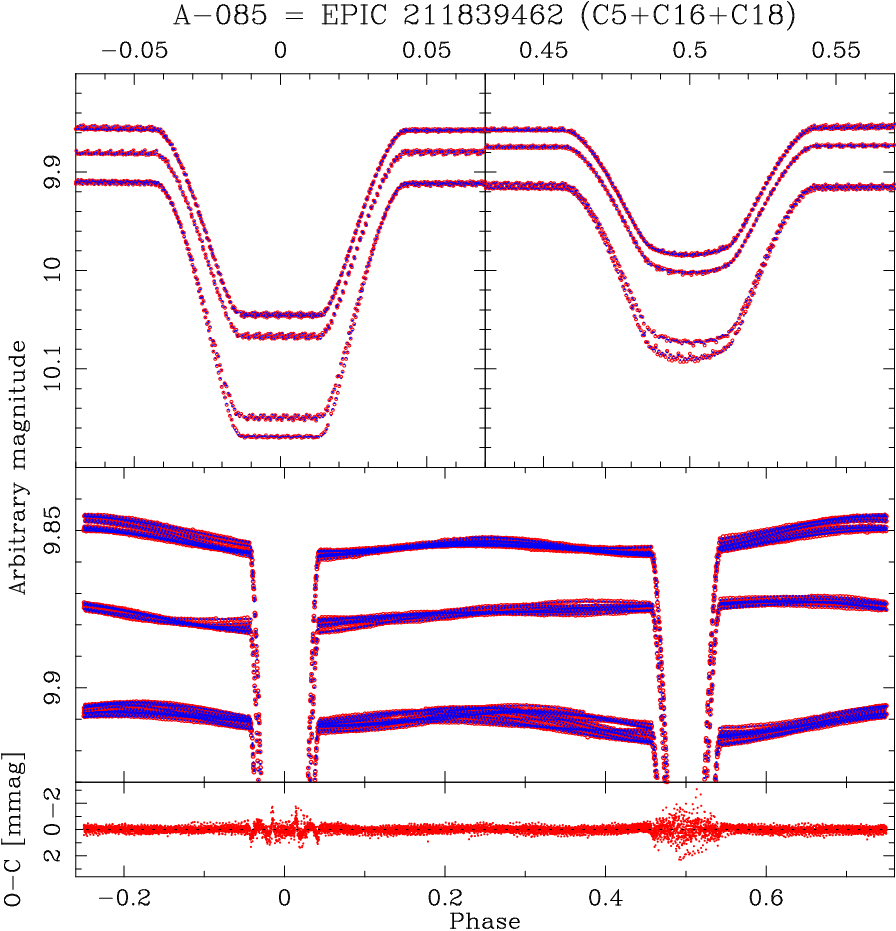}
	\includegraphics[width=0.32\textwidth]{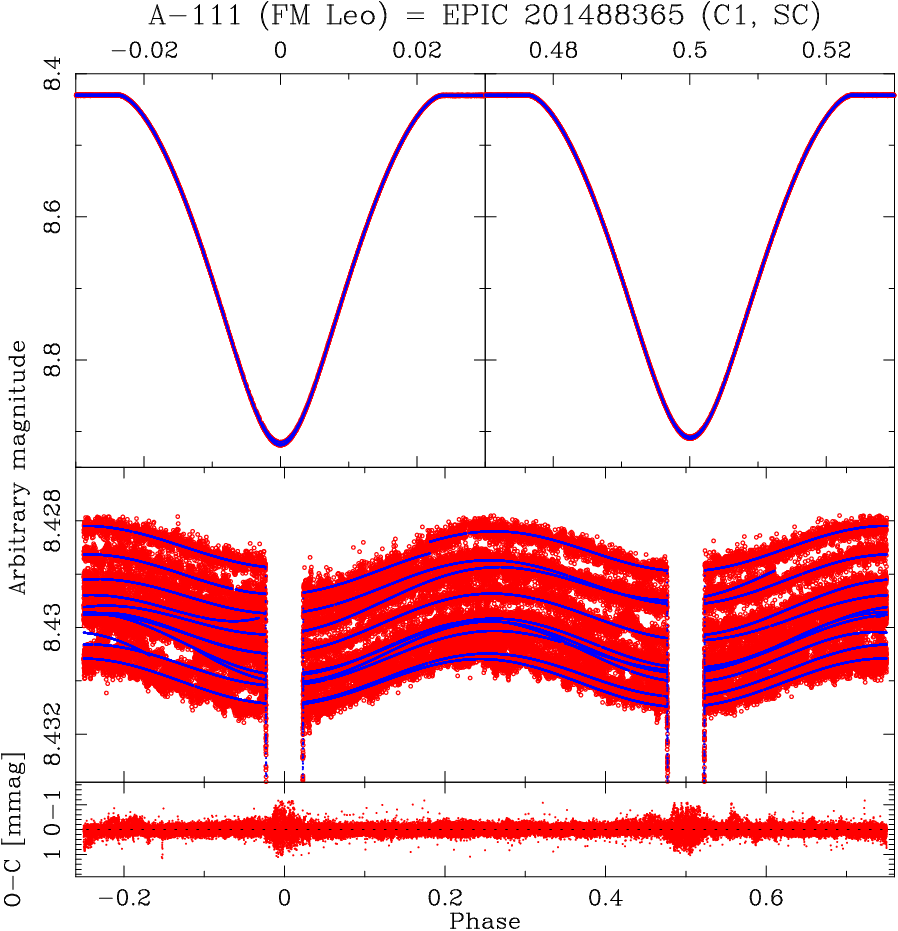}
	\includegraphics[width=0.32\textwidth]{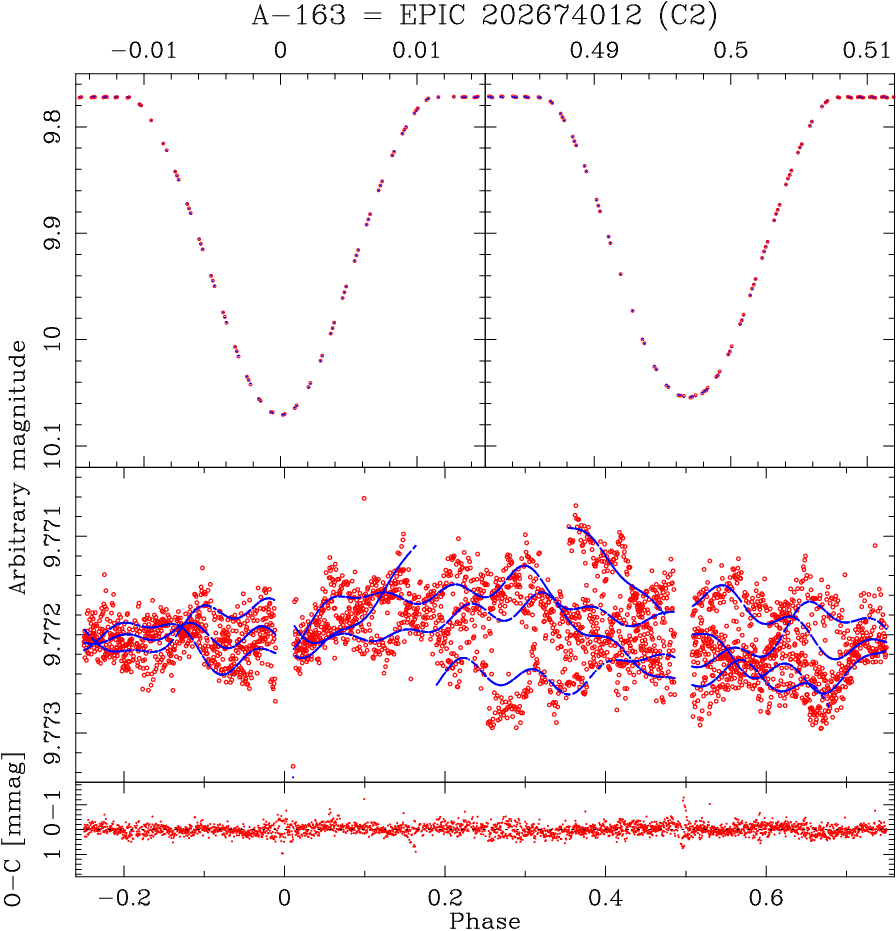}
	\includegraphics[width=0.32\textwidth]{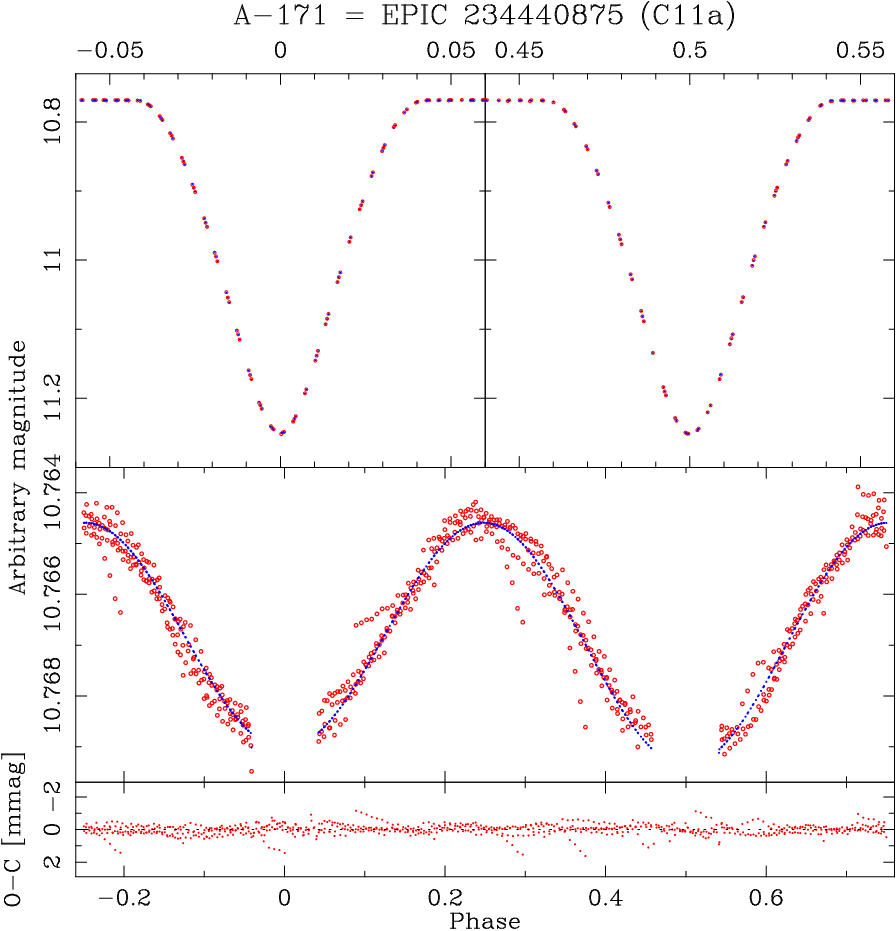}
	\includegraphics[width=0.32\textwidth]{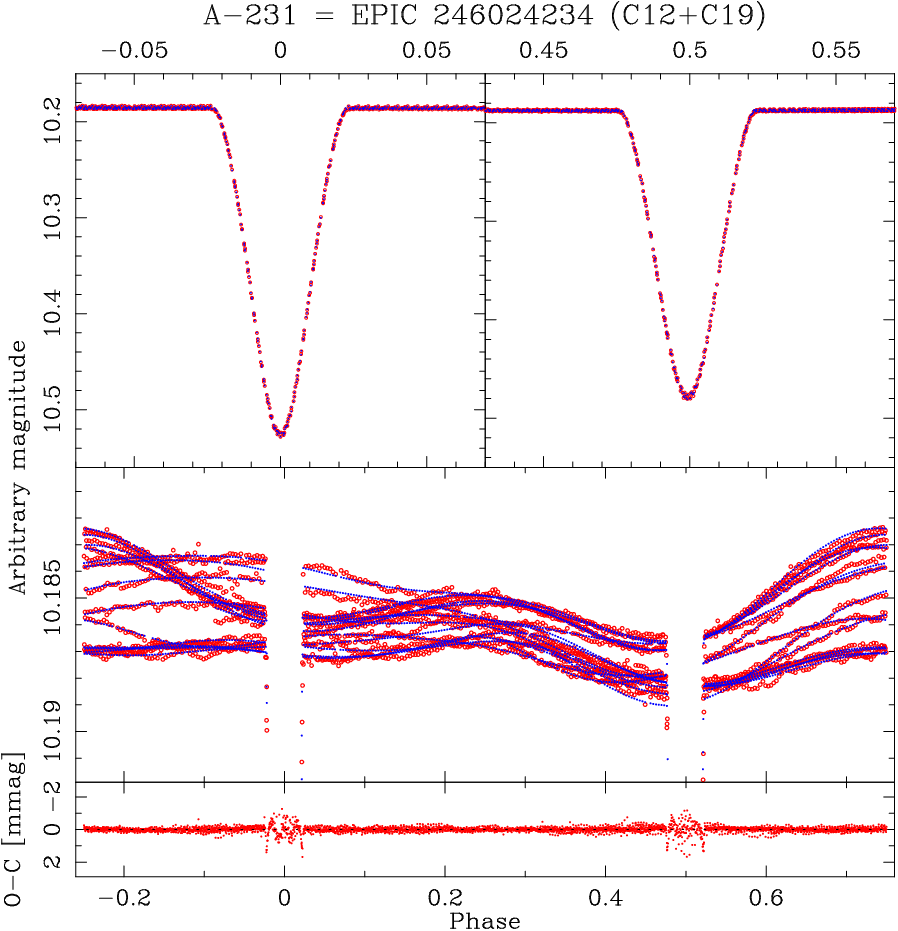}
    \caption{Observed (red) and modelled (blue) photometric \ktwo data for the studied systems, 
    phase-folded with the orbital period, with phase 0 set to the moment of the
    primary eclipse. For each system we show zooms on the primary and secondary eclipse (top), on the 
    out-of-eclipse variations (middle), and residuals of the fit. Data for all available cadences are
    shown (cadence numbers given in labels). In A-045 the scale of spot-originated variation is larger
    than the secondary eclipse. Changes in pattern of spots in time is also clearly seen for 
    A-231, A-083 and  A-085. In the last case, changes in the total brightness and eclipse 
    depths are caused by the amount of the companion's contribution varying from cadence to 
    cadence, and within C16. The short-period fluctuations in A-163 may come from oscillations.}
    \label{fig_lc1}
\end{figure*}

The secondary contributes only about 0.5\% of the total flux in the \kep photometric
band, therefore the observed very strong and rapidly evolving out-of-eclipse modulation 
(Fig.~\ref{fig_lc1}) is probably caused by spots located on the primary only. Its scale
($\sim$66~mmag) is much larger than of the ellipsoidal variations ($\sim$5~mmag), and the 
depth of the secondary eclipse ($\sim$12~mmag). The pattern of spots changes in time scales 
comparable to the short orbital period (1.653~d), meaning that every orbital revolution the 
light curves looks differently. A closer inspection shows that two brightness minima, presumably
related to two different groups of spots, move in longitude with unequal rates, which would
imply differential rotation of the primary. This obviously made the fit difficult, thus the 
resulting $rms$ of the residuals is one of the highest in the sample, and the scatter significantly 
increases during eclipses. To model this system, we split the light curve to five pieces, 
used a single 5-th order polynomial, and up to seven sine functions per piece.

Moreover, the low brightness of the secondary makes it hard to detect in the spectra. 
Its RVs were not measured in some cases. An additional obstacle was a presence of 
strong third light in the spectra, which produced a narrow peak in the cross-correlation 
function (CCF), at a position near the centre-of-mass velocity $\gamma_1$ of the system. 
The rapidly rotating primary produces a broad CCF peak, which very often interfered with 
the narrow one from the third light, making its RV measurements difficult and uncertain.
We estimate the RV of the third light to be $-28.1\pm0.4$~k\ms, very close to $\gamma_1$. 
This suggests that the narrow CCF peak may come from a star gravitationally bound to 
the eclipsing pair.

However, we have found no significant third light $l_3$ in \jkt fits. Thus, in the 
final runs we decided to hold $l_3$ fixed to 0. The reason for this is most likely the
additional flux correction adopted in the detrending scheme. Despite those 
difficulties we managed to obtain relatively good precision of 1.2-1.8\% in masses, 
and $\sim$2.2\% in radii.

In this special case the \ispec analysis was done on the primary (SNR of the recovered 
spectrum $\sim$171) and the tertiary (SNR$\sim$92). The latter was found to have 
$T_{\rm eff}$=4990(340)~K and $\log(g)$=3.16(58)~dex, which suggests it is a (sub-)giant,
therefore it is more evolved and massive than the primary. The best-fitting isochrone was found for
the age of 9.3~Gyr, and [Fe/H] of +0.05~dex, different from the \ispec value only by 
1.2$\sigma$. This isochrone predicts the tertiary to have $M\simeq1.056$~M$_\odot$ and 
$2.34<R<5.88$ R$_\odot$.

\begin{figure*}
    \centering
    \includegraphics[width=0.3\textwidth]{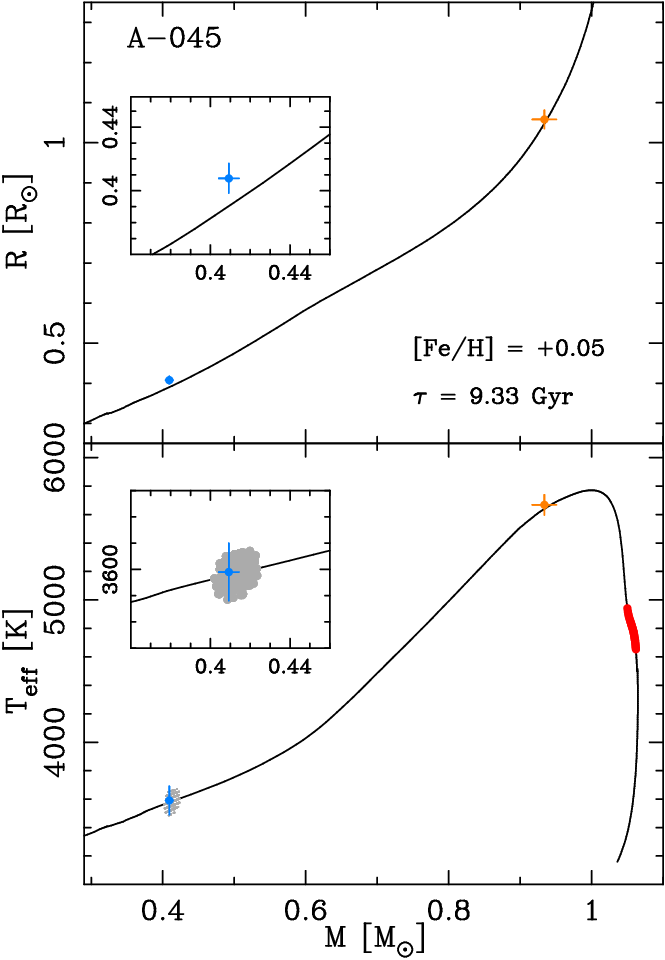}  
    \includegraphics[width=0.3\textwidth]{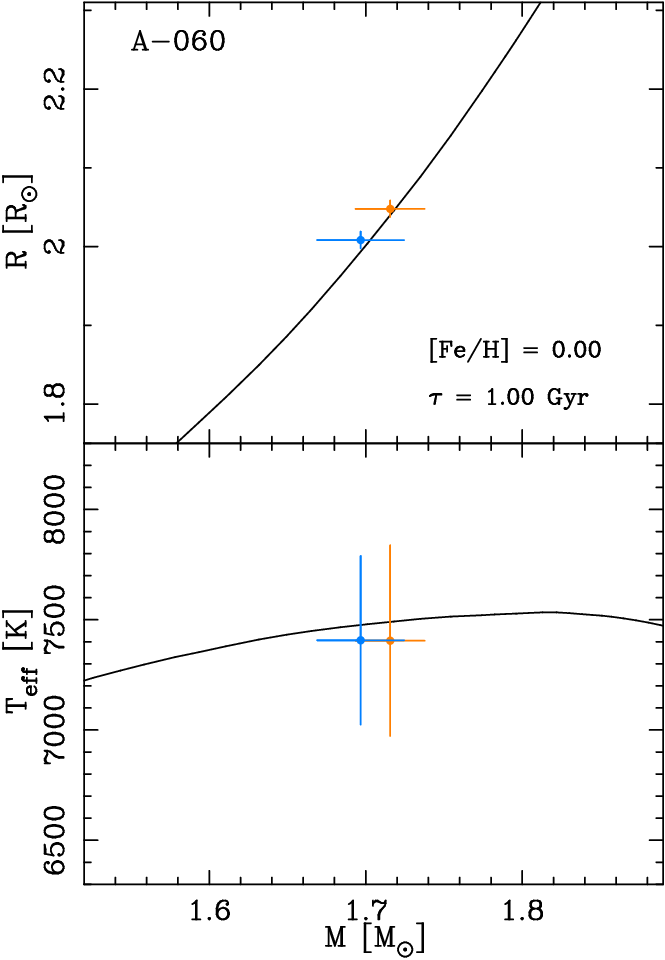}  
    \includegraphics[width=0.3\textwidth]{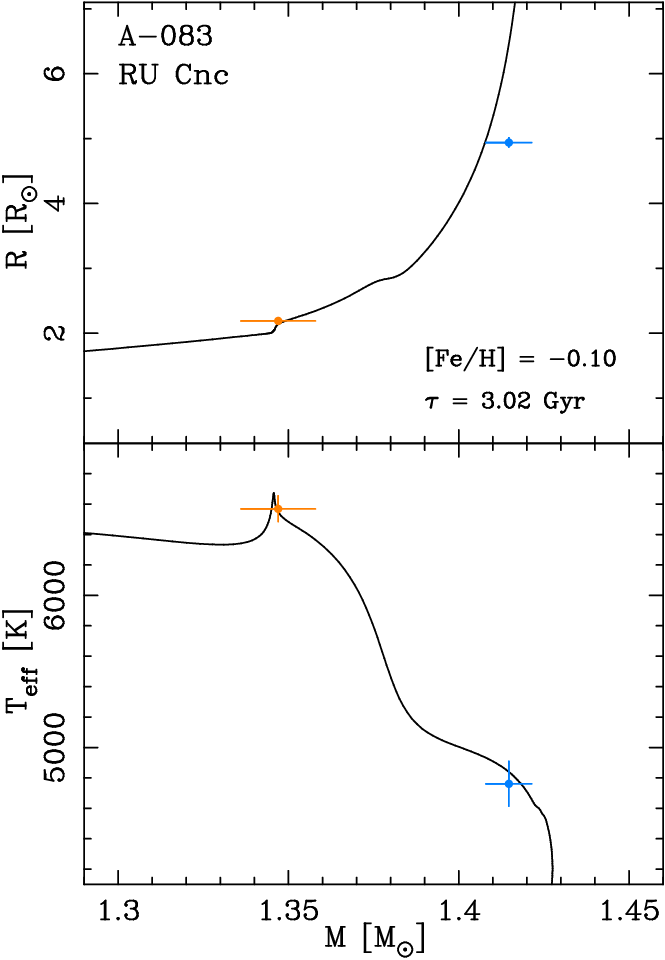}  
    \includegraphics[width=0.3\textwidth]{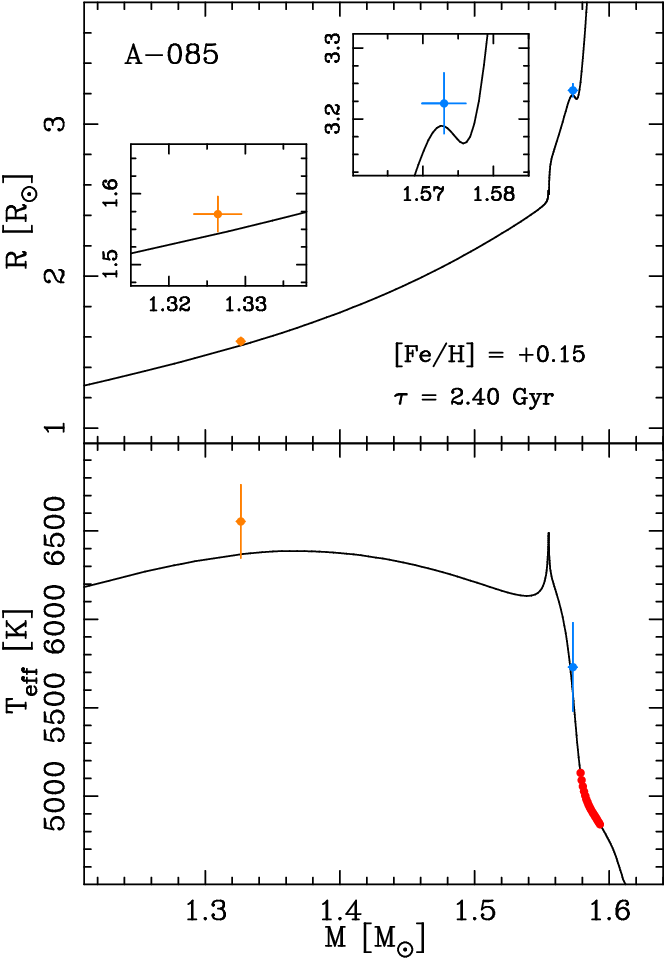}  
    \includegraphics[width=0.3\textwidth]{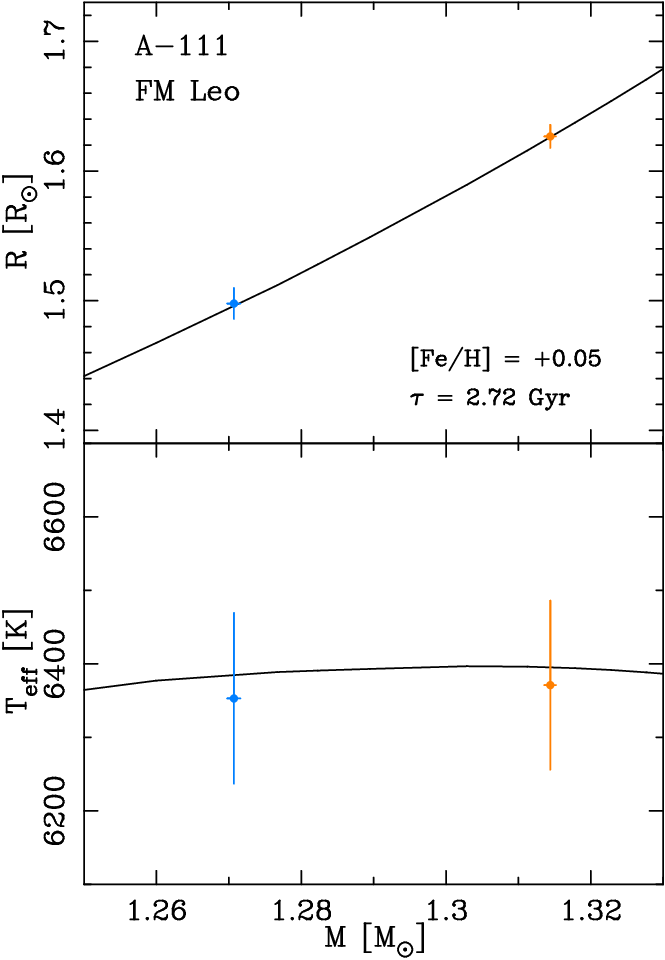}  
    \includegraphics[width=0.3\textwidth]{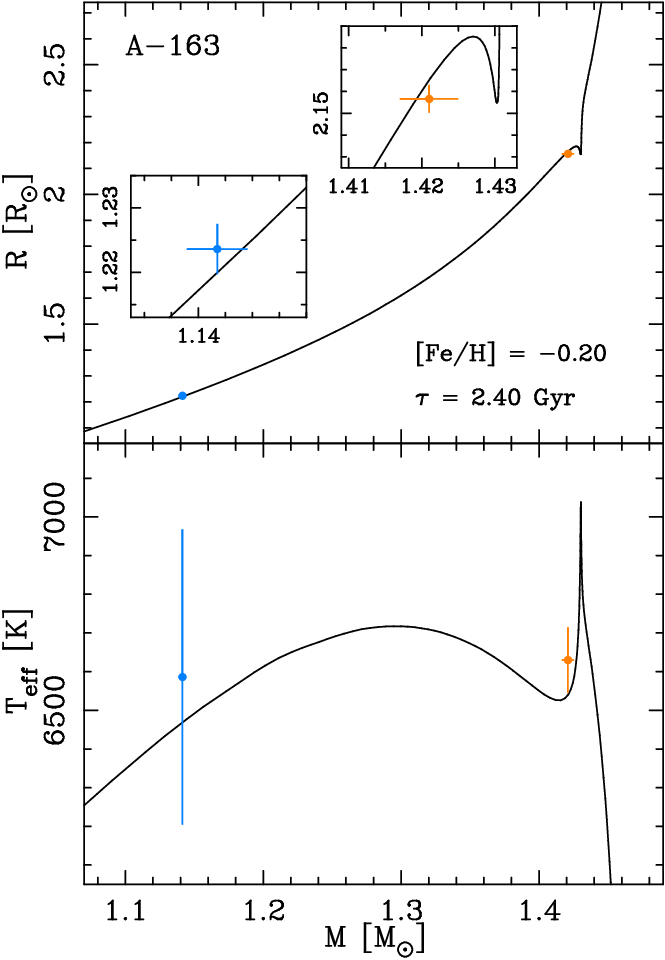}  
    \includegraphics[width=0.3\textwidth]{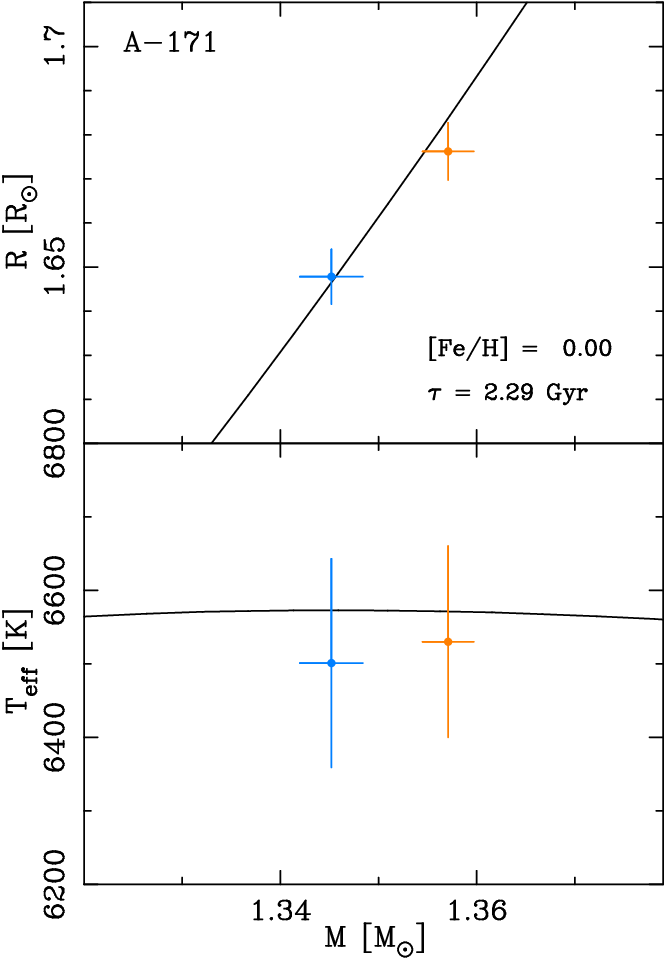}  
    \includegraphics[width=0.3\textwidth]{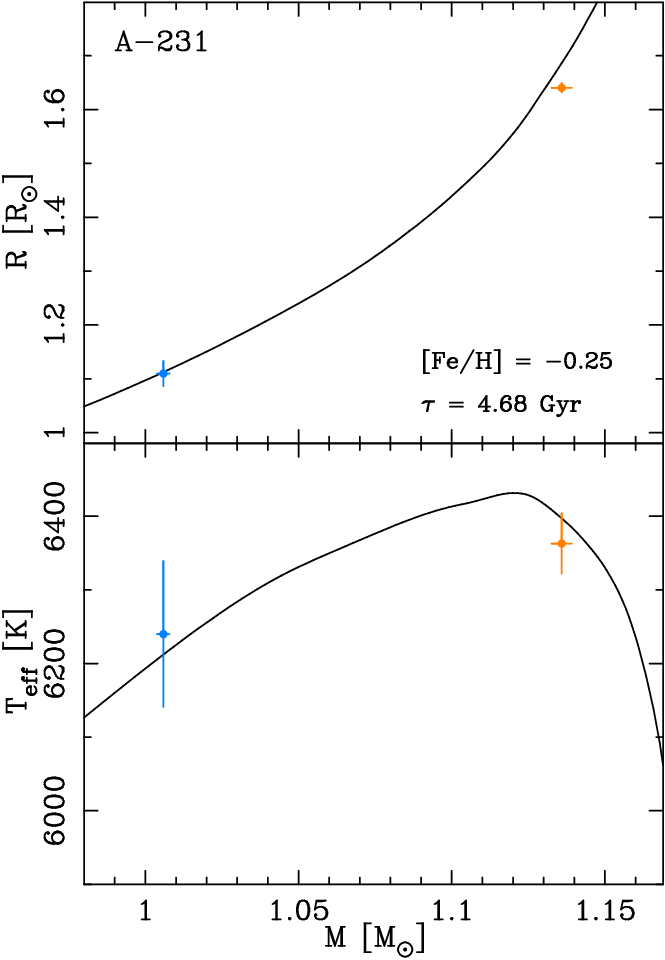}  
    \caption{Comparison of masses, radii and effective temperatures with the best-fitting isochrones. Their ages and metalicities are labelled. Orange and blue points denote primary and secondary components, respectively. Red lines mark segments of given isochrones that reproduce the measured properties ($T_{\rm eff}, \log{g}$) of tertiary companions to A-045 and A-085 (outside the panels in $M-R$ diagrams).}
    \label{fig_iso}
\end{figure*}

A-045 is the only case in our sample, where the $T_{\rm eff}$ of the secondary
was not used in isochrone fitting. Instead, it was evaluated from the distribution
of models falling within 3$\sigma$ from the other parameters. The value given in 
Table~\ref{tab_par} is calculated from this distribution. It is drawn with grey symbols
in the lower A-045 panel of Fig.~\ref{fig_iso}.

We should, however, keep in mind that for reasons discussed in Sect.~\ref{sec_iso}, 
our age estimation in this case may
be affected by the primary's activity, i.e. the radius might have been inflated, 
$T_{\rm eff}$ underestimated, and the whole system might be slightly younger than 9.3~Gyr.

\subsection{A-060}
This binary is composed of two very similar, yet not identical stars. The mass ratio $q$ is 
differs from 1 by only 1.2$\sigma$, but the fractional radii $r_1$ and $r_2$ differ by 
nearly 7$\sigma$. This is due to relatively low precision of RV measurements, 
hampered mainly by fast rotation of both components. Despite that, the precision in masses is 
still good: 1.3 and 1.7\% for the primary and secondary, respectively. 

The entire light curve (Fig.~\ref{fig_lc1}) was modelled
in one fit. Except ellipsoidal modulations ($\sim$17~mmag) and a weak long-term trend, no significant 
out-of-eclipse variations have been detected, thus only a single 5-th degree polynomial (without 
any additional sine functions) had to be used. The spread of \jkt fit residuals is rather large 
but roughly constant for every orbital phase. This allowed us to reach a very good precision 
in radii, at the level of 0.6\% for both components.

The \ispec analysis turned out pose some challenges, and resulted in large uncertainties. 
The two components were quickly found 
to be hotter than 7000~K, and rotating rapidly. This, combined with the mediocre SNR
of the disentangled spectra ($\sim$66), made the spectral features shallow. In this case,
taking into account the possible high values of temperatures, we decided to set the
$v\sin(i)$ parameter free (the calculations of $v_{\rm syn}$ implemented 
in \jktabs are relevant for cooler stars with convective envelopes). Components
of A-060 appear to rotate slower, than in case of synchronous rotation 
($\sim48.8$~k\ms).

Comparison with isochrones resulted in determination of the most probable age of 
$1.00$~Gyr for [Fe/H]=0.0~dex. Both components are on the main sequence 
(MS), and evolved from the zero-age main sequence (ZAMS). The predicted temperatures,
highest in our sample, are nearly identical, which can be deduced 
from the eclipses of almost the same depth.

An interesting feature of this system is that both components lay in the theoretical instability 
strip for $\delta$~Scuti type (dSct) pulsators. When compared to a sample of dSct pulsators in
binaries, compiled by \citet{kah17}, one can see that parameters of both components of A-060  
agree well with the sample. We can use the relations obtained by \citet{kah17} between
the period of pulsations $P_{\rm pul}$ and various stellar and binary parameters 
($P, M, R, T_{\rm eff}, \log(g), q$). The obtained values of $P_{\rm puls}$ vary from 
$\sim$0.034~d (vs. $R$), up to 0.081~d (vs. $q$), with most of the values between 
0.040 and 0.055~d. With the current data we do not detect any significant pulsation signal, 
but this may be due to the cadence of observations, and/or features of the adopted 
detrending algorithm. A better insight may come from TESS observations,
expected to be done in cycles 43 to 45.

\subsection{A-083 = RU Cnc}

\begin{figure}
    \centering
    \includegraphics[width=0.8\columnwidth]{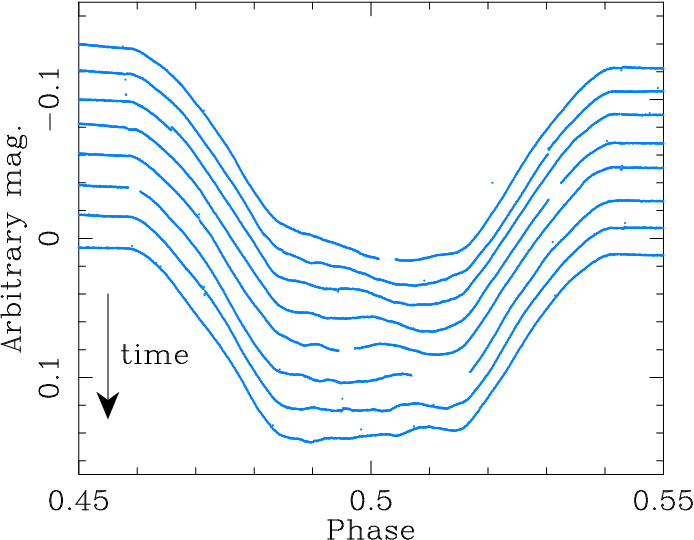}
    \caption{
    Variations in the shape of the secondary eclipse of A-083. Eight events from C5 (SC data)are shown folded in orbital phase, and arbitrarily shifted in such way, that the earliest is on the top. One can see distortions caused by the (smaller) primary component obscuring inhomogeneities on the secondary's surface. Variation in the eclipse's shape means that the location of those surface features changes from orbit to orbit.
    }
    \label{fig_rucnc_sec}
\end{figure}

RU~Cnc is a chromospherically active system of the RS~CVn type. In its \ktwo light curve 
(Fig.~\ref{fig_lc1}) we can see quickly evolving spots, flares, and a total primary
(deeper) eclipse, with some variations in its depth. These characteristics are typical
for a system with large, cold, heavily spotted component, accompanied by a smaller hot one.
The scale of the spot-originated variability ($\sim$80~mmag) overwhelms the ellipsoidal
modulation ($\sim$8~mmag).

The CCF shows two clear peaks of significantly different widths.
The RVs of the cold secondary are strongly affected by spots and rapid rotation, with the 
$rms$ of 0.90 k\ms, while that of the hot primary have significantly better
precision, with the $rms$ of 230~\ms. Both values are better than for the 
previous studies: \citet{pop90} gives 4.3 and 4.4 k\ms\ for the primary (hot) and secondary
(cold), respectively, while \citet{imb02} analogously gives 1.80 and 3.22~k\ms.
The final precision of mass determination in our solution is at a very good level 
of 0.8 and 0.5\% for the primary and secondary, respectively. The \ispec
analysis of the deconvolved spectra (SNR$\sim$86 and 58 for the primary and secondary,
respectively) suggested a small depletion of metals, but with high uncertainties.
As expected, the evolved inflated secondary turned out to be much cooler than the 
primary.

It is important to note that in his Table~2, \citet{imb02} probably confused the hot 
and cold component, and this confusion might have been taken over by other authors. 
The hotter component (spectral type F5 therein) is shown to have larger mass than 
the cooler (K1IV). Notably, the assignment in Table~3 of the same work is different.
The RV curves from Imbert's Figure~4, show that the hotter component is the one with 
more measurements (31) than the cooler K1IV-type component (21). It also seems to show 
larger RV amplitude (thus lower mass) and smaller spread ($rms$). We confirmed this by 
taking the original measurements and running an orbital fit with \vfit -- the component 
with more data points is the one with larger $K$, and smaller $rms$, which is in 
agreement with the scenario that it is less massive and slower rotating.
Table~2 shows the situation opposite to Fig.~4, where more observations and smaller 
spread were attributed to the cold component, making it the more massive one. 
Additionally, a note in the SB9 catalogue \citep{pou04} points the hotter component 
(the primary) as the one with more measurements. Our new data also clearly show that 
the hotter, earlier type primary, which produces the narrow CCF peak and lower $rms$ 
of the fit, is the less massive one (Fig.~\ref{fig_rv}, Tab.~\ref{tab_par}). 

We thus believe that Imbert unintentionally confused the components in Table~2 (but not Table~3) 
of his work, improperly making the hot one also the more massive one. It is correct at the Main 
Sequence, but that time it was already known that one of the stars is a giant. This probably led
to further interpretations
\citep[like in][]{egg17} that the later-type giant star must have undergone a substantial mass loss.
When the situation is inverted (the colder and larger companion is more massive, as in our 
solution), the inconsistency with evolutionary models vanishes (Fig.~\ref{fig_iso}), 
and no additional assumption of mass loss or other processes need to be made.

The photometric LC data consist of two Campaigns, 5 and 18, which for the \jkt analysis were 
split into 3 and 2 parts, respectively. We modelled the photometric trends
and spot-originated variations with a single 5-th degree polynomial and between five and seven sine
functions. The activity obviously affected the final results, especially $r_{1,2}$ (thus $R_{1,2}$),
whose uncertainties were dominated by the $rms$ of partial results. Still, our approach resulted in
relatively good final precision in absolute radii, at the level of 1.5-1.9\%.

A-083 is one of two systems in our sample, for which short-cadence (SC) data are available. Their
closer examination reveals multiple discontinuities, very often during the primary eclipse
(which makes it nearly impossible to properly assess its true depth), as well as high complexity 
of the spot pattern on the secondary. In Figure~\ref{fig_rucnc_sec} we show a zoom on eight 
consecutive secondary eclipses recorded in C5 SC data (detrended with {\sc everest}, obtained 
directly from MAST, and cleaned of the outliers). They reveal multiple distortion events 
which we interpret as caused by the (smaller) primary transiting in front of inhomogeneities 
on the secondary's surface. 
Observations of such phenomena in eclipsing binaries are rare, but have been reported 
\citep[e.g. KIC~10514158;][]{lur17}.

Such rapid evolution of spots in time (also seen in A-045 or A-085) makes analysis difficult 
and vulnerable to systematical errors. \citet{cok19} took the C5 long-cadence data, and found a 
solution with the {\sc phoebe} code \citep{prs05}, where properties of spots (location, size, 
temperature contrast) can be modelled, but in their model the spots are stationary, making their 
solution unrealistic (as we argue in Section~\ref{sec_lcfit}). 
Notably, \citeauthor{cok19} did not make a full fit to the C18 data, but only changed the parameters 
of spots in order to reproduce the light curve.
Additionally, their study lacks information about the details of the fitting scheme (e.g. treatment
of effective temperatures and their influence on other parameters), quality of the fit, or even properly 
presented residuals thereof, which, from the inspection of their Figure~5, seem to be quite large
and inhomogenous in phase, strongly suggesting systematical errors. Not much is said about
the error budget, and, most likely, the systematics were not properly accounted for. All this
makes the stellar parameters by \citet{cok19} unreliable, in terms of both accuracy and precision. 

In comparison, our approach allows for dynamical changes of the combined detrending and 
spot-originated brightness modulation, however without modelling the spots themselves. 
This also introduces systematical 
uncertainties (for example, the model may predict incorrect eclipse depth which affects $r_{1,2}$),
but we attempt to quantify them and incorporate into our uncertainties, as described in 
Sect.~\ref{sec_lcfit}. More accurate results are still possible to obtain through 
a dedicated, detailed analysis, that focuses on the secondary eclipses and takes 
into account the non-static character of spots, i.e. is capable of modeling the 
evolution of spot parameters during a single orbital period, including possible 
differential rotation, and very careful detrending of the SC data, so the intrinsic 
variability of short time scales and small amplitudes is not lost.
Such an effort was not in the scope of this work, and may be done in the future.

\begin{table*}
    \centering
    \caption{Comparison of our results for A-083=RU~Cnc with literature. Please note the inversion
    of masses introduced by \citeauthor{imb02}. Absolute values of $M$ and $R$ given in \citet{pop90} 
    and \citet{imb02} are estimates, and assume inclination $i=90^\circ$. Source of the 
    $T_{\rm eff}$ values in \citet{cok19} is unclear.}
    \begin{tabular}{lcccc}
    \hline
    Parameter & \citet{pop90} & \citet{imb02} (Tab.~2)$^a$ & \citet{cok19} & This work \\
    \hline \hline 
    $P$ (d)         &10.17289(-)&10.172988(2)&10.172918(3)&10.1729311(55)\\
    $K_1$ (k\ms)    & 70.4(1.2) & 67.50(71) & 68.19(8)  & 70.61(9)  \\
    $K_2$ (k\ms)    & 69.9(1.2) & 70.46(64) & 70.69(8)  & 67.2(3)   \\
    $i$ ($^\circ$)  &  90(-)    &  90(-)    & 89.7(4)   &   89(1)   \\
    $M_1$ (M$_\odot$)& 1.46(7)  &  1.42(4)  & 1.437(46) & 1.349(23) \\
    $M_2$ (M$_\odot$)& 1.47(7)  &  1.36(4)  & 1.386(44) & 1.416(12) \\
    $R_1$ (R$_\odot$)& 1.9(-)   &  1.89(2)  & 2.392(69) & 2.19(4)   \\
    $R_2$ (R$_\odot$)& 4.9(-)   &  4.83(5)  & 5.016(80) & 4.94(7)   \\
    $T_{\rm eff,1}$ (K)&  ---   &  ---      & 6860(285) & 6569(84)  \\
    $T_{\rm eff,2}$ (K)&  ---   &  ---      & 4800(200) & 4761(147)  \\
    \hline
    \end{tabular}
    \\$^a$ In Table~3 of \citet{imb02} the masses are inverted, and in agreement with this work.
    \label{tab_comp_rucnc}
\end{table*}

In Table~\ref{tab_comp_rucnc} we compare our results with those of \citet{pop90}, \citet{imb02}, 
and \citet{cok19}. Masses and radii from the two former studies come from the assumption
of $i=90^\circ$, and estimates of fractional radii from incomplete photometry. The most recent
study uses RVs of both previous authors, and probably carries on the confusion on which component
is hotter and more massive. One can see that our results are still the most precise ones, and 
probably also more accurate, while the treatment of errors is better. We are surprised to see 
that, for example, RV amplitudes $K_{1,2}$ in \citet{cok19} have errors of only 80~\ms, even though
they were derived from data of substantially worse quality ($rms$ of single k\ms) than ours.
This again undermines the reliability of these results. Our study is also the first 
where temperatures were determined in a direct way.

Comparison of our measurements with the MESA isochrones points towards the age 
of 3.02~Gyr, and metallicity below solar. The primary is at the very end of its main 
sequence evolution, while the secondary is at the red giant branch, growing
with an accelerating pace, currently about 0.02~R$_\odot$~Myr$^{-1}$. 
Using the formula from \citet{egg83}, we can estimate
the effective radius of the secondary at the moment of reaching the Roche lobe
to be 10.63~R$_\odot$. This should happen in about 110~Myr, assuming negligible loss
of mass and angular momentum, however, as \citet{cok19} have convincingly shown, the
orbital period of RU~Cnc gradually decreases as 7.9(1.2)$\times$10$^{-7}$~d~yr$^{-1}$. 
Considering this, and masses of both components, RU~Cnc will likely evolve into a 
W~UMa-type binary \citep{yil13}.

The secondary resides in a very interesting place on the $M-R$ and $T_{\rm eff}-log(g)$
diagrams,\footnote{See for example 
\url{https://www.astro.keele.ac.uk/~jkt/debcat/debplots.html}} in the so-called 
Hertzsprung gap, between old giants and sub-giants that have just 
moved from the main sequence, which is caused by a rapid growth in this phase. 
This gap is best seen on $\log(g)$ distribution, between values 3.5 and 3.0~dex.
A star of 1.415~M$_\odot$ needs about 160~Myr to go from $\log(g)=3.5$ to 3.0~dex, 
and the secondary of A-083 should reach that point in about 50~Myr.
Till now, no component of a DEB with precisely measured properties has been 
found in this region, at least among the FGK spectral types. 

RU~Cnc gives a unique opportunity to study rare stages of stellar evolution, activity, and binary
interactions but requires a special attention when comes to light curve modelling, that has 
not been applied to any other system so far. 
We have reached a good precision and accuracy in determination of stellar parameters,
but there is still space for improvement, e.g. precise element abundances are missing. 
The Community is encouraged to study this unique system, especially its chemical composition, magnetic 
field, and future tidal evolution.

\subsection{A-085}

This system is similar to the previous one: the light curve shows a total primary minimum and
out-of-eclipse variations that change in time (Fig.~\ref{fig_lc1}), the CCF is composed of a 
broad and a narrow peak, and the eclipsing pair is composed of a main sequence primary, 
and a sub-giant secondary. The main difference is that the depths of eclipses are more similar,
indicating temperature ratio closer to 1 than for RU~Cnc, and a strong and variable addition 
of the third light. This system has been observed in three campaigns -- C5, C16, and C18 --
during which the orientation of the satellite was varying. Each time the amount of flux
from the visual companion, that was falling into the pixel aperture mask, was different.
Especially, it changed drastically during 
C16, which is reflected by the change in depths of eclipses (the lowest of three A-085 curves in 
Fig.~\ref{fig_lc1}). For this reason, and also to model the varying out-of-eclipse modulation,
likely coming from spots ($\sim$10~mmag) with addition from the ellipsoidal effects 
($\sim$5~mmag), we have split the full data set in 10 parts. For each
part we used a single 5-th degree polynomial, and a number (between 5 and 7) of sine functions.
For two parts form C16 we also added a variation in $l_3$ approximated by a 3-rd degree polynomial. 

As expected from such situation, the overall error budget of light-curve-based parameters 
was dominated by their $rms$-es, which shows that we have properly taken the 
systematics into account. In case of the RVs, the activity and rotation again are the main sources 
of uncertainties for the cool secondary ($rms\simeq130$~\ms), while the earlier type primary
is more stable ($rms\simeq$90~\ms). These values are, however, among the best in our sample, 
which allowed us to reach an excellent precision in masses ($\sim$0.2\% for both primary and 
secondary), and also very good one in radii ($\sim$0.55--0.80\%),
which makes A-085 another example of a well-studied binary with an evolved, sub-giant component.

An interesting feature of A-085 is that it is part of a visual binary 
(ADS~7030~AB, $\rho\simeq9$~arc-sec), 
and the whole system is a triple. The companion is separated far enough to obtain its
spectra independently, therefore the information about the companion can be used to further 
constrain the age and metallicity of the whole system, or to verify results obtained 
for the eclipsing pair. Additionally, we also took advantage of the total part of
the primary eclipse to obtain useful information \citep[see ][for another example]{hel15}. 
In total, the spectral analysis in \ispec was performed on five spectra: disentangled primary,
disentangled secondary, totality secondary (all from CHIRON), tertiary from HDS, and 
tertiary from CORALIE (shift-and-stack of three visits). Their SNR values were 68, 46, 40, 
103, and 50, respectively. All five contributed to the final [M/H], and the last two to
constrain the age and verify the solution. The totality spectrum also allowed to verify 
if the \todcor flux ratios and further renormalisation had produced reliable results.
The averaged results for the tertiary's parameter are $T_{\rm eff}$=4994(161)~K, and
$\log(g)$=3.39(28)~dex, making it a sub-giant as well.

Our values of $M, R, T_{\rm eff}, l_2/l_1$ and [M/H] are very well reproduced by a 
$\tau=2.40$~Gyr, [Fe/H]=+0.15~dex isochrone. The primary is at the main sequence, while 
the secondary evolved to a sub-giant, and is currently growing and cooling down,
although not as quickly as the secondary in RU~Cnc. 

The isochrone reproduces the tertiary's atmospheric
properties for $M\simeq1.587$~M$_\odot$, and a wide range of radii: $3.27<R<5.23$~R$_\odot$.
The {\it Gaia Data Relase 2} \citep[GDR2;][]{gai18} gives the effective temperature of ADS~7030~B at 
$4992^{+84}_{-52}$~K, and the estimated radius is $4.78^{+0.10}_{-0.15}$~R$_\odot$. 
The latest LAMOST data release No.~5 \citep[LDR5;][]{luo19}, 
gives a temperature 5012(15)~K (average of two entries) and $\log(g)=3.359(25)$~dex, with
metallicity slightly above solar -- 0.085(15)~dex. All those values agree reasonably
well with our estimates, and the 2.40~Gyr, 0.15~dex isochrone. Finally, the observed 
2MASS $J-K$ colour index 0.56(2)~mag is also very well reproduced by $\tau=2.40$~Gyr, 
[Fe/H]=+0.15~dex around 1.59~M$_\odot$ (0.53-0.59~mag). We therefore 
conclude that our solution for A-085 is consistent, evolutionary status of the whole triple 
is well established, and the visual companion B is likely a $\sim$1.59~M$_\odot$, 
$\sim$5000~K sub-giant.

\subsection{A-111 = FM Leo}

\begin{table*}
    \centering
    \caption{Comparison of our results for A-111=FM~Leo with literature. Values common for two
    works are given between their respective columns, and are originally derived in the earlier work.}
    \begin{tabular}{lccccc}
    \hline
    Parameter & \citet{rat10} & \citet{max18} & \citet{syb18} & \citet{gra21} & This work \\
    \hline \hline 
    $P$ (d)         &6.728606(6)&6.728609(2)&6.7286134(36)&6.7286133(8)&6.728606(2)\\
    $K_1$ (k\ms)    & \multicolumn{2}{c}{76.62(27)} & 75.992(29)& 76.017(10)& 76.033(19)\\
    $K_2$ (k\ms)    & \multicolumn{2}{c}{78.46(28)} & 78.653(36)& 78.654(15)& 78.646(22)\\
    $i$ ($^\circ$)  & 87.98(6)  & 87.96(1)  & 89.07(63) &87.941(23) & 87.939(28)\\
    $M_1$ (M$_\odot$)& 1.318(7) &  1.32(1)  & 1.3119(16)& 1.3144(5) & 1.3144(8) \\
    $M_2$ (M$_\odot$)& 1.287(7) &  1.29(1)  & 1.2675(14)& 1.2703(4) & 1.2707(7) \\
    $R_1$ (R$_\odot$)& 1.648(43)&  1.634(5) & 1.76(8)   & 1.625(2)  & 1.627(9)  \\
    $R_2$ (R$_\odot$)& 1.511(49)&  1.498(6) & 1.22(11)  & 1.508(3)  & 1.498(12) \\
$T_{\rm eff,1}$ (K) & 6316(240) & 6430(155) &  ---      &  6397(56) & 6371(115) \\    
$T_{\rm eff,2}$ (K) & 6190(211) & 6420(155) &  ---      &  6386(56) & 6353(116) \\    
    \hline
    \end{tabular}
    \label{tab_comp_fmleo}
\end{table*}

A-111 = FM Leo is the brightest target in our sample. It has been recognized
by \citet{max18} as a potentially very useful system for testing evolutionary models, since
its brightness and low level of activity allow for very precise and 
accurate photometry and RVs. Unsurprisingly, it is listed in the DEBCat catalogue.
Indeed, among our systems, A-111 has the lowest $rms$ of both RV and light curves 
(Tab.~\ref{tab_par}). The \ktwo data show relatively low-scale out-of-eclipse modulations, 
mainly ellipsoidal variations ($\sim$1~mmag) and long-time-scale trends (Fig.~\ref{fig_lc1})
Some obvious remnants from detrending also are noticeable, mainly discontinuities, 
or short-period oscillations. We split the SC light curve into five 
parts, and in the modelling we used a single 5-th degree polynomial and a single 
sine function, except for the first part 
where the sine was not used, and the fourth part, when two polys were used.
The \ispec analysis of disentagled spectra (SNR$\sim$117 and 95 for the primary
and secondary, respectively) showed a small metal depletion, and very similar 
temperatures of the components.

We reached excellent precision in masses ($\sim$0.06\% for both components), 
and also very good one in radii (0.55 and 0.8\%), confirming the conclusions of 
\citet{max18}. This is thanks to the superb data, which include the richest
set of high-precision RVs used for this object to date.

We found an excellent agreement of our results with a [Fe/H]=0.05~dex, $\tau=2.72$~Gyr
isochrone. Both components are therefore on the main sequence. Comparison of our results 
with previously published works is shown in Table~\ref{tab_comp_fmleo}. 
The agreement is very good, except for $R_2$ from \citet{syb18}, who only used
the ASAS light curve, with no additional photometry. In particular we confirm the
values of radii found by \citet{max18} from \ktwo long cadence light curve extracted
with {\sc k2sff}. The advantage of using better precision RV measurements 
\citep[][and this work vs. previous studies]{syb18,gra21} is clearly seen. The 
better precision reached by \citet{gra21} is most likely a result of their choice of 
the WD code to model the system, and to model the curve as a whole. 
This code, however, may be underestimating the systematic uncertainties, which could be seen 
in the case of AI~Phe in the thorough study of systematics and reliability in DEB 
modelling by \citet{max20}. 

Notably, \citet{gra21} reported detection of a Doppler beaming effect coming from only one of 
the components. Following the same approach, which uses the formalism presented in 
\citet{pla19}, one can conclude that both components of FM~Leo should produce 
indistinguishable signals (dependent on $T_{\rm eff}$ and $\log(g)$), that would 
cancel out. We suspect that the asymmetry found in their out-of-eclipse variations 
(long cadence data in these orbital phases) is caused by weak, cold spots.

Nevertheless, we conclude that FM~Leo can be a real benchmark binary for 
testing stellar structure and evolution models.

\subsection{A-163}

\begin{table*}
    \centering
    \caption{Comparison of our results for A-163 with literature. Values common for two
    works are given between their respective columns, and are originally derived in the earlier work.}
    \begin{tabular}{lcccc}
    \hline
    Parameter & \citet{max18} & \citet{hel18} & \citet{hoy20} & This work \\
    \hline \hline 
    $P$ (d)         &\multicolumn{2}{c}{23.30962(5)}& 23.309637(9) & 23.309595(35) \\
    $K_1$ (k\ms)    & 45.3(2.4) & 45.34(17) & 45.8(3) & 45.44(4)  \\
    $K_2$ (k\ms)    & 57.8(2.6) & 56.35(21)$^a$ & 55.1(7) & 56.56(7) \\
    $e$             & 0.027(2)  & 0.0272(13)& 0.004(1) & 0.026(2)  \\
    $\omega$ ($^\circ$) & 259.7(6)  & 259(3) & 263.1(1.4) & 259(1)    \\
    $i$ ($^\circ$)  &\multicolumn{2}{c}{88.65(1)}& 88.6(17)$^b$ & 88.668(14)  \\
    $M_1$ (M$_\odot$)& 1.48(16) & 1.407(12) & 1.35(5) & 1.421(4) \\
    $M_2$ (M$_\odot$)& 1.16(13) & 1.132(9)  & 1.12(6) & 1.141(2) \\
    $R_1$ (R$_\odot$)& 2.18(7)  & 2.154(7)  & 1.63(5) & 2.157(6) \\
    $R_2$ (R$_\odot$)& 1.22(4)  & 1.209(6)  & 1.18(5) & 1.224(4)  \\
$T_{\rm eff,1}$ (K) &\multicolumn{2}{c}{6250(285)}& 6550(150) & 6630(83) \\    
$T_{\rm eff,2}$ (K) &\multicolumn{2}{c}{6150(285)}& 6275(250) & 6586(380) \\    
    \hline
    \end{tabular}
    \label{tab_comp_a163}
  \\$^a$ The value of $K_2$ in Table~1 of \citet{hel18} is incorrectly given as 53.35 k\ms, which 
    is a typographical error. The value given here is correct. Other parameters in \citet{hel18},
    including those dependent on $K_2$ are given correctly.
  \\$^b$ Original notation from Table~2 in \citet{hoy20} with a probable typographic error.
\end{table*}

This is a system with the longest orbital period in our sample. The time span of the \ktwo light 
curve is only $\sim$3 times the orbital period, therefore a more precise determination of $P$
came from the RVs. The LC curve was fitted in \jkt as a whole, but two different 4-th degree
polynomials were used (for data taken before and after JD=2456937.0), together with five sine
functions (Fig.~\ref{fig_lc1}). Because there are
remnant oscillations seen in the residuals, we used the RS approach in \jkt to assess realistic
errors. Still, the resulting $rms$ of the light curve fit is one of the lowest in our sample.

The RV data ($rms$ of 100 and 220 \ms for the primary and secondary, respectively) allow for 
excellent precision in mass determination, namely 0.28 and 0.18\% for the primary and secondary,
respectively. Coupled with high precision photometry, our data set led to a very good precision
also in radii: $\sim$0.3\% for both components. The \ispec analysis of deconvolved spectra
(SNR$\sim$127 and 38 for the primary and secondary, respectively) points towards sub-solar 
metallicity stars of similar temperatures, but different $\log(g)$. As we did not assume 
synchronous rotation, we fitted for $v\sin(i)$ and obtained 14.1(7) and 5.7(2.8)
k\ms, for the primary and secondary, respectively. The primary is therefore rotating
faster than in (pseudo-)synchronous configuration (4.68~k\ms). Its rotation may also
be seen in the light curve -- the sine with the largest amplitude was identified 
with $P=8.334~d$ (Tab.~\ref{tab_pierdy}), which corresponds to $v\sin(i)\simeq13.2$~k\ms. 
The secondary also seems to rotate super-synchronously (2.66~k\ms in tidal equilibrium), 
but the uncertainty is larger. According to \jktabs results, synchronisation should 
occur at the age of 
$\sim$3.7~Gyr, which we can treat as an upper limit for the age of A-163.

In Table~\ref{tab_comp_a163} we compare our results
with previous works: by (1) \citet{max18} who used only four publicly available FEROS 
spectra and calculated RVs with their own approach, (2) \citet{hel18}, where many more spectra
were used, but only the orbital solution has been updated with respect to the former,
and (3) by \citet{hoy20} who used a mixture of literature and their own RV measurements, 
own light curve solution, and spectral analysis.  

The improvement in precision of absolute parameters with respect to the earliest results is clear, 
coming mainly from the improved spectroscopic orbit. The precision in radii we obtained in this 
paper is therefore slightly better than in \citet{hel18}, although the errors in $r_{1,2}$ 
themselves \citep[which are the same in][]{max18, hel18} are a bit worse. 
This is possibly because \citet{max18},
did not take into account all possible sources of systematic errors, especially given the low number 
of eclipses in the \ktwo data, or the influence of the third light, which we found small but not
negligible. In any case, our independent analysis of a different RV set and with different approach 
to the light curve, led to high-quality results, consistent with the two previous studies.

Notably, the stellar parameters from the most recent study of \citet{hoy20} are not only of 
lower precision, but also show disagreement with other studies, mainly when comes to the primary 
component. This might have been caused by several factors, e.g. fixing $l_2/l_1$ to a value 
obtained from the $V$ band, instead of {\it Kepler}'s, incorrect BJD calculated for the FEROS and 
HARPS spectra, higher measurement errors than in our approach to the same data, or obsolete version 
of the \jkt code used for light curve modeling. It is also not explained which \ktwo data product 
was used, or was the light curve produced with a different software. We have also found some 
internal inconsistencies in \citet{hoy20}. For example, taking the RVs as they are given in their 
Table~A1 leads to $rms$ of 1.4-1.7 k\ms, instead of 0.41-0.52 cited in Table~2. We could also not 
reproduce their results of $K_1$ and $K_2$, obtaining 45.3 and 54.1 k\ms, 
respectively\footnote{For this test, we used our \vfit code and the measurements listed by 
\citet{hoy20}.}, which is more than 1$\sigma$ off the values given by them. Finally, applying their 
partial orbital, light curve, and spectral analysis results to \jktabs we could not reproduce 
the cited errors, nor the distance of 257(11)~pc. The same apparent magnitudes (in $B,V,J,H$, 
and $K$, from {\it Simbad}), effective temperatures
(from their spectral analysis) and fractional radii lead to $d\simeq$210~pc (using surface 
brightness-$T_{\rm eff}$ relations from \citeauthor{ker04}, \citeyear{ker04}). Finally,
their \ispec results suggest the secondary rotates faster (16.3~k\ms) than the primary
(14.7~k\ms), while no such situation is observed in the width of the CCF peaks. For all these 
reasons we find the solution of \citet{hoy20} questionable, and suggest a cautious approach 
to their results for A-163 and other objects in their paper.

The MESA isochrone that best reproduces our results was found for the age of 
2.40~Gyr and metallicity of -0.20~dex. The age is under the upper limit we obtained from
the predicted synchronicity age, supporting the super-synchronous rotation of the primary.
It also suggests a spin-orbit misalignment in this binary. 
The primary is somewhat evolved and is about 
to leave the main sequence. Interestingly, it lays at the edge of the $\gamma$~Dor 
theoretical instability strip, and its properties ($T_{\rm eff}, \log(g), v_{\rm mic}$) 
fall into their observed distributions \citep{kah16}. As was mentioned before, our model
with 5 sines still does not cover all of the frequencies found in the data. One of the 
stronger and better separated remnants is a peak at $P = 0.98$~d, which would be in
agreement with the $v\sin(i) - \log(P_{\rm puls})$ relation shown in \citet{kah16}.

We therefore conclude that the primary of A-163 may be a $\gamma$~Dor-type star,
but a more detailed analysis of its light curve is necessary, which was not the scope
of this work. We would like to encourage the Community to pursue further investigation,
by noticing that the TESS 30-min data are available from two sectors (12, 39), and the
out-of-eclipse brightness variations are visible (with the 8.3~d peak being the dominant one).
The measurement of the RV effect would also be welcome, as it could confirm the spin-orbit
misalignment.

\subsection{A-171}
This system shows very little out-of-eclipse variations, except for ellipsoidal 
($\sim$4~mmag) and imperfect 
detrending effects (Fig.~\ref{fig_lc1}). In particular, low level of activity allowed us to obtain 
relatively good $rms$ of the light curve fit (0.33~mag) and RV fit for both components (0.54 and 
0.26~k\ms, despite notable rotational broadening of $\sim$27~k\ms). The LC data set was fitted as
a whole, with no additional polynomials or sine functions needed, but in this case it was 
necessary to free the reflection coefficients, and they were found to be 2.75(0.12)$\times10^{-3}$
and 2.58(0.11)$\times10^{-3}$ for the primary and secondary, respectively. 
Furthermore, the \ktwo light curve covers $\sim$5.5 orbital periods, so the analysis probably 
did not suffer much from systematics that might have originated from low number of data points 
and insufficient sampling.

We found that A-171 is composed of two very similar, but not identical F-type dwarfs. 
We reached a very good relative precision in masses and radii, 
namely $\sim$0.3\% in $M$ and 0.4\% in $R$. The spectroscopic analysis of the 
disentangled spectra (both of SNR$\sim$70) led to almost 
identical temperatures, which was expected from the nearly equal depths of eclipses.
The metallicity was found to be solar within errors.

We note that both components of A-171 have similar mass to the primary of RU~Cnc, but are 
significantly smaller and hotter. This, suggests that A-171 is younger than RU~Cnc. 
The MESA isochrones nicely reproduce our results with the $\tau=2.29$~Gyr, 
[Fe/H]=0.00~dex isochrone. Both components are thus on the Main Sequence.

\subsection{A-231}\label{sec_a231}
A-231 is another system with prominent spots that change rapidly, and thus the shape of its
light curve varies from orbit to orbit (Fig.~\ref{fig_lc1}). However, in contrast to A-083 and 
A-085, the eclipses are partial, suggesting that the difference between two 
radii is not that significant as in the two aforementioned cases. The spot-induced 
brightness modulation is much weaker, not exceeding 5~mmag, yet larger than the ellipsoidal
effect ($\sim$1~mmag). The \ktwo data from campaign C12 were split in 3
parts, and for each of them a single 5-th degree polynomial and five sine functions were used 
in the fitting. The data from campaign C19 were fitted with the aid of one 5-th degree poly and 
2 sines, but they cover only a single orbital period. Remnant, short-time-scale brightness 
modulations are still visible in the residuals. Their origin is unclear to us, possibly a 
mixture of systematics coming from 
the detrending and real, intrinsic oscillations. As in the case of A-163, for error
calculations in \jkt we used the RS, which gives more reliable results in such cases. 
Nevertheless the overall quality of the fits was 
still very good, with the final $rms$ of residuals of the fit at the level of 0.19~mmag 
(0.11~mmag outside of eclipses). The most important parameters ($r_{\rm 1,2}, J, l_2/l_1, i$) 
obtained from various parts of the light curve turned out to be quite stable.

The activity did not significantly hamper the RV fit. With the $rms$ of only 82 and 
147 \ms for the primary and secondary, respectively, we were able to reach precision 
in masses at the level of 0.19--0.27\%. At the same time, the precision in radii was 
several times lower: 0.5 and 2.1\% for the primary and secondary, respectively.

The \ispec analysis of the disentangled spectra (SNR$\sim$167 and 75 for the
primary and secondary, respectively) revealed that the system is most likely 
metal-depleted, and the components have quite similar temperatures -- the primary
had cooled significantly down during its MS evolution. Notably, this is the only case
where $[\alpha/Fe]$ was found different from 0.0 by more than 1$\sigma$ -- 0.10(7)~dex.

Our estimates are very well reproduced by a $\tau=4.68$~Gyr,
[Fe/H]=-0.25~dex isochrone. A-231 is therefore the second oldest, and the
most metal-depleted system in our sample. The metal depletion explains 
why the secondary is significantly larger than the Sun, despite having nearly the same 
mass and similar age.

\section{Conclusions}\label{sec_conc}

We presented results of a combined RV + photometric + spectral analysis of eight 
detached eclipsing binaries from our spectroscopic survey, that were observed 
during the \ktwo mission. Special attention was put to model the variable 
out-of-eclipse modulations, and properly account them for in the error budget. 

The use of high-precision photometric data, together with good quality RV measurements, allowed 
us to obtain valuable results, especially absolute masses and radii, with fractional errors
below 3\%, and very often below 1\%. The presented systems show interesting and rare 
properties, such as: low mass (secondary of A-045), multiplicity (A-045, A-085),
possible pulsations and spin-orbit misalignment (A-163), later
stages of evolution (A-085), including the first object crossing the Hertzsprung gap with 
$M$ and $R$ measured with high precision (A-083 = RU~Cnc), or potential for obtaining results of 
extremely high precision of 0.1\% and better (A-111 = FM~Leo), which is important for testing 
the latest models of stellar evolution \citep{val17}. The Community is encouraged to pursue 
further, more detailed studies of the systems presented here.

We conclude that space-borne photometry opens possibility for extremely precise $<$0.5\%
derivation of stellar radii, but only for systems that systems do not show prominent 
and unstable spots.
Otherwise the precision is largely affected by their evolution in time which is likely
disturbed by the detrending algorithms, and therefore has to be properly
treated in error estimations. We claim that it is still possible to reach $\sim$0.1\% level
of radii uncertainty for highly-spotted DEBs, but light curve modelling of such 
data should include the dynamical character of the spots characteristics.

\section*{Acknowledgements}
We would like to thank Dr. C. Ga\l an from the NCAC Warsaw for sharing with us his results
of spectral analysis of FM~Leo before the publication, and the anonymous Referees for the 
comments and corrections to the manuscript.
This research made use of data collected at ESO under programmes 082.D-0499, 084.B-0029, 
086.D-0078, 089.D-0097, 091.D-0145, 091.D-0414, 095.D-0026, 100.D-0273, 100.D-0339, 
101.D-0697, and 102.D-0281 as well as 
through CNTAC proposals CN2011B-021, CN2012A-021, CN2012B-036, CN2013B-022, CN2014A-044, 
and CN2015A-074.
This research is based in part on data collected at Subaru Telescope, which is operated by the National Astronomical Observatory of Japan. We are honored and grateful for the opportunity of observing the Universe from Maunakea, which has the cultural, historical and natural significance in Hawaii.
This work has made use of data from the European Space Agency (ESA) mission
{\it Gaia} (\url{https://www.cosmos.esa.int/gaia}), processed by the {\it Gaia}
Data Processing and Analysis Consortium (DPAC,
\url{https://www.cosmos.esa.int/web/gaia/dpac/consortium}). Funding for the DPAC
has been provided by national institutions, in particular the institutions
participating in the {\it Gaia} Multilateral Agreement.
K.G.H. is supported by the Polish National Science Center through grant no. 2016/21/B/ST9/01613.
A.M., T.P., and M.K. are supported by the Polish National Science Center through grant no. 2017/27/B/ST9/02727.
N.E. would like to thank the Gruber Foundation for its generous support to this research. %
A.J. and R.B. acknowledge support from ANID - Millennium Science Initiative - ICN12\_009.
A.J. acknowledges additional support from FONDECYT project 1210718.
This work was supported by the Poznan University of Technology under the grant No. 0211/SBAD/0121, and by JSPS KAKENHI Grant Number 16H01106.

\section*{Data availability}
The data underlying this article are available in the article and in its online supplementary material.





\begin{thebibliography}{99}
\bibitem[Armstrong et al.(2015)]{arm15} Armstrong D. J., et al. 2015, A\&A, 579, A19 
\bibitem[Barros et al.(2016)]{bar16} Barros S. C. C., Demangeon O., Deleuil M., 2016, A\&A, 594, A100
\bibitem[Basri \& Shah(2020)]{bas20} Basri G., Shah R., 2020, ApJ, 901, 14
\bibitem[Bessell et al.(1998)]{bes98} Bessell M. S., Castelli F., Plez B., 1998, A\&A, 333, 231
\bibitem[Blanco-Cuaresma et al.(2014)]{bla14} Blanco-Cuaresma S., Soubiran C., Heiter U., Jofr\'e P., 2014, A\&A, 569, A111
\bibitem[Blanco-Cuaresma et al.(2016)]{bla16} Blanco-Cuaresma, S., et al. 2016, in 19th Cambridge Workshop on Cool Stars, Stellar Systems, and the Sun (CS19), 22
\bibitem[Brahm et al.(2017)]{bra17} Brahm R., Jord\'an A., \& Espinoza N. 2017, PASP, 129, 034002
\bibitem[Cardelli et al.(1989)]{car89} Cardelli J.~A., Clayton G.~C., Mathis J.~S., 1989, ApJ, 345, 245
\bibitem[Ceraski(1911)]{cer11} Ceraski W., 1911, AN, 187, 77
\bibitem[Choi et al.(2016)]{cho16} Choi J., Dotter A., Conroy C., Cantiello M., Paxton B., Johnson B.~D., 2016, ApJ, 823, 102
\bibitem[Clausen et al.(2008)]{cla08} Clausen J. V. et al., 2008, A\&A, 487, 1095
\bibitem[Code et al.(1976)]{cod76} Code A. D., Bless R. C., Davis J., Brown R. H., 1976, ApJ, 203, 417
\bibitem[\c{C}okluk et al.(2019)]{cok19} \c{C}okluk K. A., Ko\c{c}ak D., I\c{c}li T., Karak\"{o}se S., \"{U}st\"{u}nda\v{g} S., Yakut K., 2019, MNRAS, 488, 4520
\bibitem[Conroy et al.(2020)]{con20} Conroy K. E., et al., 2020, ApJS, 250, 34
\bibitem[Coronado et al.(2015)]{cor15} Coronado J., et al., 2015, MNRAS, 448, 1937
\bibitem[Cutri et al.(2003)]{cut03} Cutri R.~M., et al., 2003, yCat, II/246
\bibitem[Devor et al.(2008)]{dev08} Devor J., Charbonneau D., O'Donovan F. T., Mandushev G., Torres G., 2008, AJ, 135, 850
\bibitem[Dotter(2016)]{dot16} Dotter A., 2016, ApJS, 222, 8
\bibitem[Eggleton(1983)]{egg83} Eggleton P. P., 1983, ApJ, 268, 368
\bibitem[Eggleton \& Yakut(2017)]{egg17} Eggleton P. P., Yakut K., 2017, MNRAS, 468, 3533
\bibitem[Flower(1996)]{flo96} Flower P. J., 1996, ApJ, 469, 335
\bibitem[Gaia Collaboration(2016)]{gai16} Gaia Collaboration, 2016, A\&A 595, A1
\bibitem[Gaia Collaboration(2018)]{gai18} Gaia Collaboration, 2018, A\&A, 616, A1
\bibitem[Gaia Collaboration(2021)]{gai21} Gaia Collaboration, 2021, A\&A, 649, A1 
\bibitem[Gillen et al.(2014)]{gil14} Gillen E., et al., 2014, A\&A, 562, A50
\bibitem[Gillen et al.(2017)]{gil17} Gillen E., et al., 2017, ApJ, 849, 11 
\bibitem[Girardi et al.(2002)]{gir02} Girardi L., et al., 2002, A\&A, 391, 195
\bibitem[Graczyk et al.(2021)]{gra21} Graczyk D., et al., 2021, A\&A, 649, A109
\bibitem[Gray \& Corbally(1994)]{gra94} Gray R. O., Corbally C. J., 1994, AJ, 107, 742
\bibitem[Grevesse et al.(2007)]{gre07} Grevesse N., Asplund M., Sauval A. J., 2007, Space Sci. Rev., 130, 105
\bibitem[Gustafsson et al.(2008)]{gus08} Gustafsson B., Edvardsson B., Eriksson K., J\o rgensen U. G., Nordlund \AA., Plez B., 2008, A\&A, 486, 951
\bibitem[He{\l}miniak et al.(2014)]{hel14} He{\l}miniak K.~G., Brahm R., Ratajczak M., Espinoza N., Jord{\'a}n A., Konacki M., Rabus M., 2014, A\&A, 567, A64
\bibitem[He{\l}miniak et al.(2018)]{hel18} He{\l}miniak K. G., Jord\'{a}n A., Espinoza N., Brahm R., Konacki M., 2018, RNAAS, 2, 226 
\bibitem[He{\l}miniak et al.(2019c)]{hel19c} He{\l}miniak K.~G., Jord{\'a}n A., Espinoza N., Brahm R., 2019c, 
Proc. of the International Astronomical Union, 15(S354), 300-304
\bibitem[He\l miniak et al.(2011)]{hel11} He\l miniak K.~G., et al., 2011, A\&A, 527, A14
\bibitem[He\l miniak et al.(2012)]{hel12} He\l miniak K. G., et al., 2012, MNRAS, 425, 1245
\bibitem[He\l miniak et al.(2015)]{hel15} He\l miniak K. G., et al., 2015, MNRAS, 448, 1945
\bibitem[He\l miniak et al.(2016)]{hel16} He\l miniak K. G., et al., 2016, MNRAS, 461, 2896
\bibitem[He{\l}miniak et al.(2019a)]{hel19} He\l miniak K.~G., et al., 2019a, A\&A, 622, A114
\bibitem[He{\l}miniak et al.(2019b)]{hel19b} He\l miniak K.~G., et al., 2019b, MNRAS, 484, 451
\bibitem[Hippke et al.(2019)]{hip19} Hippke M., David T.~J., Mulders G.~D., Heller R., 2019, AJ, 158, 143
\bibitem[Huber et al.(2016)]{hub16}  Huber D., et al., 2016, ApJS, 224, 2
\bibitem[Ho et al.(2017)]{ho17} Ho A.~Y.~Q., et al., 2017, ApJ, 836, 5 
\bibitem[Howell et al.(2014)]{how14} Howell S. B., et al., 2014, PASP, 126, 398
\bibitem[Hoyman \& \c{C}ak\i rl\i(2020)]{hoy20} Hoyman B., \c{C}ak\i rl\i\, \"O., 2020, MNRAS, 493, 2329 
\bibitem[Iliji\'c et al.(2004)]{ili04} Iliji\'c S., Hensberge H., Pavlovski K., Freyhammer L.~M., 2004, ASPC, 318, 111
\bibitem[Imbert(2002)]{imb02} Imbert M., 2002, A\&A, 387, 850
\bibitem[Ioannidis \& Schmitt(2016)]{ioa16} Ioannidis P., Schmitt J. H. M. M., 2016, A\&A, 594, A41
\bibitem[Izumiura(1999)]{izu99} Izumiura H., 1999, in: Proc. 4th East Asian Meeting on Astronomy, ed. P. S. Chen, Kunming, Yunnan Observatory, p. 77
\bibitem[Jord\'an et al.(2014)]{jor14} Jord\'an A., et al., 2014, AJ, 148, 29
\bibitem[Kahraman Ali\c{c}avu\c{s} et al.(2016)]{kah16} Kahraman Ali\c{c}avu\c{s} F., et al., 2016, MNRAS, 458, 2307
\bibitem[Kahraman Ali\c{c}avu\c{s} et al.(2017)]{kah17}Kahraman Ali\c{c}avu\c{s} F., Soydugan E., Smalley B., Kub{\'a}t J., 2017, MNRAS, 470, 915
\bibitem[Kambe~et~al.(2013)]{kam13} Kambe E., et al., 2013, PASJ, 65, 15
\bibitem[Kaufer et al.(1999)]{kau99} Kaufer A., et al. 1999, The Messenger, 95, 8
\bibitem[Kazarovets et al.(1999)]{kaz99} Kazarovets A. V., Samus N. N., Durlevich O. V., Frolov M. S., Antipin s. V., Kireeva N. N., Pastukhova E. N., 1999, IBVS, 4659, 1
\bibitem[Kervella et al.(2004)]{ker04} Kervella P., Th\'evenin F., Di~Folco E., S\'egransan D., 2004, A\&A, 426, 297
\bibitem[Kiraga \& St\k{e}pie\'{n}(2013)]{kir13} Kiraga M., St\k{e}pien K., 2013, AcA, 63, 53
\bibitem[Klinglesmith \& Sobieski(1970)]{kli70} Klinglesmith D. A., Sobieski S., 1970, AJ, 75, 175
\bibitem[Konacki et al.(2010)]{kon10} Konacki M., Muterspaugh M. W., Kulkarni S. R., He\l miniak,~K.~G., 2010, ApJ, 719, 1293
\bibitem[Kordopatis et al.(2013)]{kor13} Kordopatis G., et al. 2013, AJ, 146, 134
\bibitem[Kurucz(1992)]{kur92} Kurucz R. L., 1992, in Barbury B., Renzini A., eds, Proc. IAU Symp. 149, The Stellar Population of Galaxies, Kluwer Academic Publishers, Dordrecht, p. 225
\bibitem[Kunder et al.(2017)]{kun17} Kunder A., et al. 2017, AJ, 153, 75
\bibitem[Lanza et al.(2004)]{lan04} Lanza A.~F., Rodon{\`o} M., Pagano I., 2004, A\&A, 425, 707
\bibitem[LaCourse et al.(2015)]{lac15} LaCourse D. M., et al. 2015, MNRAS, 452, 3561 
\bibitem[Lastennet \& Valls-Gabaud(2002)]{las02} Lastennet E., Valls-Gabaud D., 2002, A\&A, 396, 551 \bibitem[Lee(2015)]{lee15} Lee C.-S., 2015, MNRAS, 453, 3474
\bibitem[Lucy(1967)]{luc67} Lucy L.~B., 1967, ZA, 65, 89
\bibitem[Luger et al.(2016)]{lug16} Luger R., Agol E., Kruse E., Barnes R., Becker A., Foreman-Mackey D., Deming D., 2016, AJ, 152, 100
\bibitem[Luger et al.(2018)]{lug18} Luger R., Kruse E., Foreman-Mackey D., Agol E., Saunders N., 2018, AJ, 156, 99
\bibitem[Luo et al.(2019)]{luo19} Luo A.-L., et al., 2019, yCat, V/164
\bibitem[Lurie et al.(2017)]{lur17} Lurie J. C., et al., 2017, AJ, 154, 250
\bibitem[Marcadon et al.(2020)]{mar20} Marcadon F., et al., 2020, MNRAS, 499, 3019
\bibitem[Mason et al.(2001)]{mas01} Mason B. D., Wycoff G. L., Hartkopff W. I., Douglass G. G., Worley C. E., 2001, AJ, 122, 6 
\bibitem[Maxted(2018)]{max18a} Maxted P.~F.~L., 2018, RNAAS, 2, 39
\bibitem[Maxted \& Hutcheon(2018)]{max18} Maxted P.~F.~L., Hutcheon R.~J., 2018, A\&A, 616, A38
\bibitem[Maxted et al.(2020)]{max20} Maxted P.~F.~L., et al., 2020, MNRAS, 498, 332
\bibitem[Moultaka et al.(2004)]{mou04} Moultaka J., Ilovaisky S.~A., Prugniel P., Soubiran C., 2004, PASP 116, 693
\bibitem[Noguchi et al.(2002)]{nog02} Noguchi K., et al., 2002, PASJ, 54, 855
\bibitem[Paredes et al.(2021)]{par21} Paredes L. A., et al., 2021, AJ, 162, 167
\bibitem[Paxton et al.(2011)]{pax11} Paxton B., Bildsten L., Dotter A., Herwig F., Lesaffre P., Timmes F., 2011, ApJS, 192, 3
\bibitem[Paxton et al.(2013)]{pax13} Paxton B., et al., 2013, ApJS, 208, 4
\bibitem[Paxton et al.(2015)]{pax15} Paxton B., et al., 2015, ApJS, 220, 15
\bibitem[Paxton et al.(2018)]{pax18} Paxton B., et al., 2018, ApJS, 234, 34
\bibitem[Pepper et al.(2008)]{pep08} Pepper J., et al., 2008, AJ, 135, 907
\bibitem[Petigura et al.(2015)]{pet15} Petigura E. A., et al., 2015, ApJ, 811, 102
\bibitem[Placek(2019)]{pla19} Placek B., 2019, Journal of Physics Conference Series, 1239, 012008
\bibitem[Pojma\'{n}ski(2002)]{poj02} Pojma\'{n}ski G., 2002, AcA, 52, 397
\bibitem[Popper(1990)]{pop90} Popper D. M., 1990, AJ, 100, 247
\bibitem[Popper \& Etzel(1981)]{pop81} Popper D. M., Etzel P. B., 1981, AJ, 86, 102
\bibitem[Pourbaix et al.(2004)]{pou04} Pourbaix D., et al. 2004, A\&A, 424, 727
\bibitem[Pr\v{s}a \& Zwitter(2005)]{prs05} Pr\v{s}a A., Zwitter T., 2005, ApJ, 628, 426
\bibitem[Ratajczak et al.(2010)]{rat10} Ratajczak M., et al. 2010, MNRAS, 402, 2424
\bibitem[Ratajczak et al.(2013)]{rat13} Ratajczak M., He{\l}miniak K.~G., Konacki M., Jord{\'a}n A., 2013, MNRAS, 433, 2357
\bibitem[Ratajczak et al.(2021)]{rat20} Ratajczak M., et al. 2021, MNRAS, 500, 4972
\bibitem[Rucinski(1969)]{ruc69} Rucinski S. M., 1969, AcA, 19, 24
\bibitem[Schwab et al.(2012)]{sch12} Schwab Ch., Spronck J., Tokovinin A., Szymkowiak A., Giguere M., Fisher D., 2012, Proc. SPIE, 8446, paper 8446-9
\bibitem[Shapley(1913)]{sha13} Shapley H., 1913, ApJ, 38, 158
\bibitem[Smith et al.(2021)]{smi21} Smith G. D., et al., 2021, arXiv:2109.00836
\bibitem[Somers et al.(2020)]{som20} Somers G., Cao L., Pinsonneault M.~H., 2020, ApJ, 891, 29
\bibitem[Southworth(2015)]{sou15} Southworth J., 2015, ASPC, 496, 164
\bibitem[Southworth et al.(2004a)]{sou04a} Southworth J., Maxted P. F. L., Smalley B., 2004a, MNRAS, 351, 1277
\bibitem[Southworth et al.(2004b)]{sou04b} Southworth J., Zucker S., Maxted P. F. L., Smalley B., 2004b, MNRAS, 355, 986
\bibitem[Southworth et al.(2011)]{sou11} Southworth J., Pavlovski K., Tamajo E., Smalley B., West~R.~G., Anderson D. R., 2011, MNRAS, 414, 3740
\bibitem[Stumpe at al.(2012)]{stu12} Stumpe M. C., et al., 2012, PASP, 124, 985
\bibitem[Sybilski et al.(2018)]{syb18} Sybilski P., Paw{\l}aszek R. K., Sybilska A., Konacki M., He{\l}miniak K. G., Koz{\l}owski S. K., Ratajczak M., 2018, MNRAS, 478, 1942
\bibitem[Ting et al.(2018)]{tin18} Ting Y.-S., Hawkins K., Rix H.-W., 2018, ApJL, 858, L7
\bibitem[Tokovinin et al.(2013)]{tok13} Tokovinin A., et al., 2013, PASP, 125, 1336 
\bibitem[Torres et al.(2010)]{tor10} Torres G., Andersen J., Gim\'{e}nez A., 2010, A\&ARv, 18, 67\bibitem[Valle et al.(2017)]{val17} Valle G., Dell'Omodarme M., Prada Moroni P.~G., Degl'Innocenti S., 2017, A\&A, 600, A41
\bibitem[Vanderburg \& Johnson(2014)]{van14} Vanderburg A., Johnson J. A., 2014, PASP, 126, 948
\bibitem[Voges et al.(1999)]{vog99} Voges W., et al. 1999, A\&A, 349, 389
\bibitem[Wilson \& Devinney(1971)]{wil71} Wilson R. E., Devinney E. J., 1971, ApJ, 166, 605
\bibitem[Wilson(2012)]{wil12} Wilson R. E., 2012, AJ, 144, 73
\bibitem[Windmiller et al.(2010)]{win10} Windmiller G., Orosz J.~A., Etzel P.~B., 2010, ApJ, 712, 1003
\bibitem[Wraight et al.(2011)]{wra11} Wraight K. T., White G. J., Bewsher D., Norton A. J., 2011, MNRAS, 416, 2477
\bibitem[Y{\i}ld{\i}z(2014)]{yil13} Y{\i}ld{\i}z M., 2014, MNRAS, 437, 185
\bibitem[Zucker \& Mazeh(1994)]{zuc94} Zucker S., Mazeh T., 1994, ApJ, 420, 806

\end{thebibliography}




\appendix
\section{Individual RV measurements}

\begin{table*}
\centering
\caption{All RVs used in this work for orbital solutions. Data for A-163 overlap with 
\citet{hel18}.
The complete table is available as supplementary material.}\label{tab_RV}
\begin{tabular}{lcccrcl}
\hline \hline
BJD & $v_1$ & $v_2$ & Target & $t_{\rm exp}$ & SNR & Inst.\\
-2450000 & (k\ms ) &  (k\ms )  & & (s) & \\
\hline
7021.117441 &  -81.04$\pm$0.89	&  --- 		  	    & A-045	&  900	&  62 	& OH \\
7024.212351 &  -85.73$\pm$2.76	&  114.31$\pm$4.19	& A-045	& 2000	&  60	& OH \\
7059.095898 &  -87.85$\pm$1.02	&  108.50$\pm$4.27	& A-045	& 1800	&  36	& OH \\
...\\
6695.616152 &  131.66$\pm$1.63	&  -90.29$\pm$0.72	& A-060	&  900	&  43 	& CHf \\
6707.544032 &  -17.95$\pm$0.83	&   60.28$\pm$0.70	& A-060	&  900	&  48	& CHf \\
6719.554731 &  -56.96$\pm$1.33	&  101.15$\pm$0.90	& A-060	&  900	&  35	& CHf \\
...\\
\hline
\end{tabular}
\end{table*}

In Table~\ref{tab_RV} we present individual RV measurements used
in this work for orbital solutions. For both components of a given binary we show the measured 
RV values $v_{1,2}$, their errors $\epsilon_{1,2}$ (both in k\ms), as well as 
exposure times in seconds and SNR at $\lambda\sim5500$~\AA. The last column
(``Inst.'') codes the telescope, spectrograph, and observing mode used:
OH = OAO-188/HIDES, MF = MPG-2.2m/FEROS, EC = Euler/CORALIE, 
CHf and CHs = SMARTS 1.5m/CHIRON fiber and slicer, EHc and EHg = ESO-3.6m/HARPS
ECHE and EGGS, OE = OHP-1.93/ELODIE.

\section{\ktwo light curves}

\begin{table*}
    \centering
    \caption{\ktwo photometric data used in this work, together with the resulting orbital phase and \jkt model.
    The magnitude scale is arbitrary and may not coincide with the observed magnitude in the \kep band.
    Data for A-111 = FM~Leo (EPIC~201488365) were taken in short cadence. Only a portion is shown here, the full Table is available as supplementary material.}
    \label{tab_alldata}
    \begin{tabular}{lrrrrcc}
    \hline
    BJD-245000 & Magnitude & Phase & Model & $(O-C)$ & EPIC~ID & Camp.\\
    \hline \hline
    7820.581500 &  9.983082 & 0.86788862 &  9.984936 & -0.001854 & 247605441 & 13 \\
    7820.601932 &  9.984303 & 0.88024685 &  9.985845 & -0.001543 & 247605441 & 13 \\
    7820.622364 &  9.985404 & 0.89260495 &  9.986778 & -0.001374 & 247605441 & 13 \\
    ... & \\
    6770.433229 & 12.203283 & 0.09490711 & 12.203177 &  0.000106 & 202073040 & 0 \\
    6770.453661 & 12.202893 & 0.10453924 & 12.202028 &  0.000865 & 202073040 & 0 \\
    6770.474092 & 12.201474 & 0.11417089 & 12.200844 &  0.000630 & 202073040 & 0 \\
    ... & \\
    \hline
    \end{tabular}
\end{table*}

\begin{figure*}
    \centering
    \includegraphics[width=0.97\textwidth]{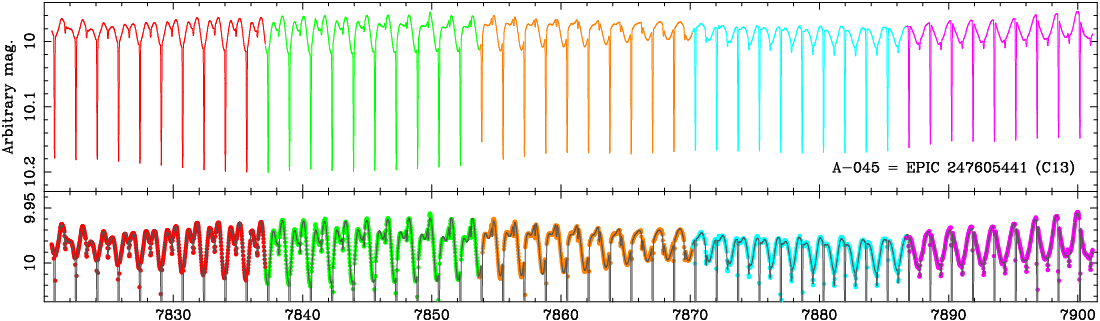} 
    \includegraphics[width=0.97\textwidth]{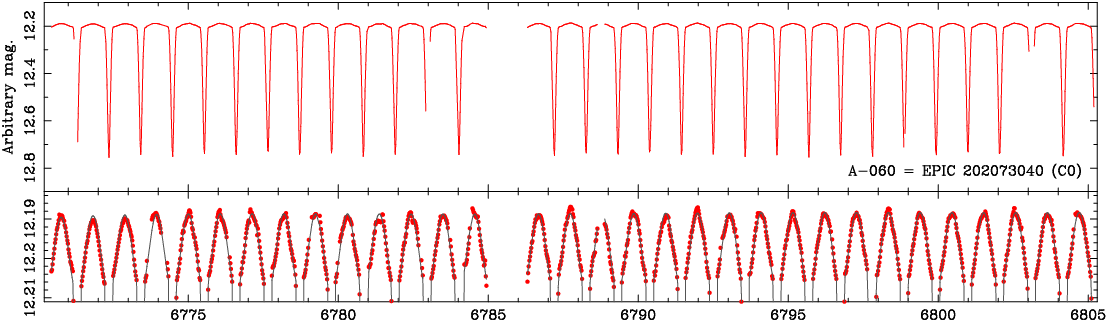}  
    \caption{The \ktwo light curves used in this work, shown as a function of time (BJD-2450000)
    and for each target and campaign separately. When analysis was done in parts, they are 
    shown with different colours. The magnitude scale for A-085 is constant in all campaigns.
    Lower panels show zooms on the out-of-eclipse modulations, in order to better present the models 
    (grey lines) and variations of the light curve shapes in time. The complete figure is
    available as supplementary material.}
    \label{fig_lcsum}
\end{figure*}

In Table~\ref{tab_alldata} and Figure~\ref{fig_lcsum} we present all the photometric data used in 
this study, as well as the model values (grey line in the figure) and residuals (in Table only).
In the figure, each target and campaign is shown separately.
If data from a single campaign were divided into parts, each part 
is drawn with a different colour. There may be systematic offsets in magnitude zero points, 
and differences between catalogue value of brightness in the \kep band and maximum value 
on the plots may occur. This does not, however, affect the results of the LC analysis.

In Table~\ref{tab_pierdy} we list the values of polynomial coefficients and properties of 
the sine functions that were used to model the out-of-eclipse modulations.


\begin{table*}
\centering
\caption{Information on polynomials and sines used in every individual \jkt model.
The polynomial is of the form $c_0 + c_1 x + c_2 x^2 +...$
The values in ``Beginning-End'', ``Pivot$_i$'', ``Range$_i$'', ad $sT_i$ (sine's time base) 
are given as BJD-2450000. Sine's periods ($sP_i$) and amplitudes ($sA_i$) are given 
in days (d) and magnitudes (mag), respectively. 
The complete Table is available as supplementary material. \label{tab_pierdy}}
\scriptsize
\begin{tabular}{ccccccc}
\hline \hline
\multicolumn{7}{c}{\it Running name and EPIC ID} \\
Campaign-Part  & $sT_1$& $sP_1$ & $sA_1$ & $sT_2$& $sP_2$ & $sA_2$ \\
Beginning-End  & $sT_3$& $sP_3$ & $sA_3$ & $sT_4$& $sP_4$ & $sA_4$ \\
            & $sT_5$& $sP_5$ & $sA_5$ & $sT_6$& $sP_6$ & $sA_6$ \\
            & $sT_7$& $sP_7$ & $sA_7$ \\
        & \multicolumn{2}{c}{Range$_1$} & Pivot$_1$ \\
        & $c_{01}$ & $c_{11}$ & $c_{21}$ & $c_{31}$ & $c_{41}$ & $c_{51}$ \\
        & \multicolumn{2}{c}{Range$_2$} & Pivot$_2$ \\
        & $c_{02}$ & $c_{12}$ & $c_{22}$ & $c_{32}$ & $c_{42}$ & $c_{52}$ \\
\hline
\multicolumn{7}{c}{\it A-045 = EPIC 247605441} \\
C13-1           & 7816.0969347285 & 1.6631766916 & 0.0128635335 & 7820.2936623448 & 0.8192959629 & 0.0094314870 \\
7820.58-7837.12 & 7821.0807661721 & 0.8353737272 &-0.0055340900 & 7821.1813222412 & 1.5723085304 &-0.0036088888 \\
                & 7820.7277959660 & 1.2677163863 &-0.0009682236 & 7821.6855947399 & 0.4077588724 & 0.0006512112 \\
                & 7820.8481968310 & 0.5391985203 & 0.0004684519 \\
    & \multicolumn{2}{c}{7820.58-7837.12} & 7828.000      \\
    & 0.0045582502 &-0.0005812301 &-0.0000828304 & 0.0000324034 & 0.0000009727 & 0.0000002941 \\
    &\multicolumn{2}{c}{---} & ---      \\
    & --- & --- & --- & --- & --- & --- \\
C13-2           & 7839.3858457234 & 1.6425099802 & 0.0178860261 & 7840.2389735631 & 1.7936060255 & 0.0014213383 \\
7837.12-7853.65 & 7840.3172866189 & 0.8193793759 &-0.0121861995 & 7842.4763761296 & 0.4066475451 & 0.0005301067 \\
                & 7840.3683741844 & 0.8571783149 &-0.0006999417 & 7842.3732345035 & 0.7934518107 & 0.0013810449 \\
                & 7842.0974614240 & 0.5463785489 &-0.0007092010 \\
    & \multicolumn{2}{c}{7837.12-7853.65} & 7846.000      \\
    & 0.0050098338 &-0.0001075007 & 0.0000359708 & 0.0000042225 &-0.0000002770 &-0.0000000447 \\
        &\multicolumn{2}{c}{---} & ---      \\
        & --- & --- & --- & --- & --- & --- \\
... \\
\multicolumn{7}{c}{\it A-060 = EPIC 202073040} \\
C0-1            & --- & --- & --- & --- & --- & --- \\
6770.43-6805.19 & --- & --- & --- & --- & --- & --- \\
                & --- & --- & --- & --- & --- & --- \\
                & --- & --- & --- \\
    & \multicolumn{2}{c}{6770.43-6805.19} & 6790.000      \\
    &-0.0009548237 &-0.0000443111 &-0.0000006063 & 0.0000003812 & 0.0000000017 &-0.0000000011 \\
    &\multicolumn{2}{c}{---} & ---      \\
    & --- & --- & --- & --- & --- & --- \\
... \\
\hline
\end{tabular}{}
\end{table*}

\section{Testing the methodology on synthetic data}
\begin{figure*}
    \centering
    \includegraphics[height=4.2cm]{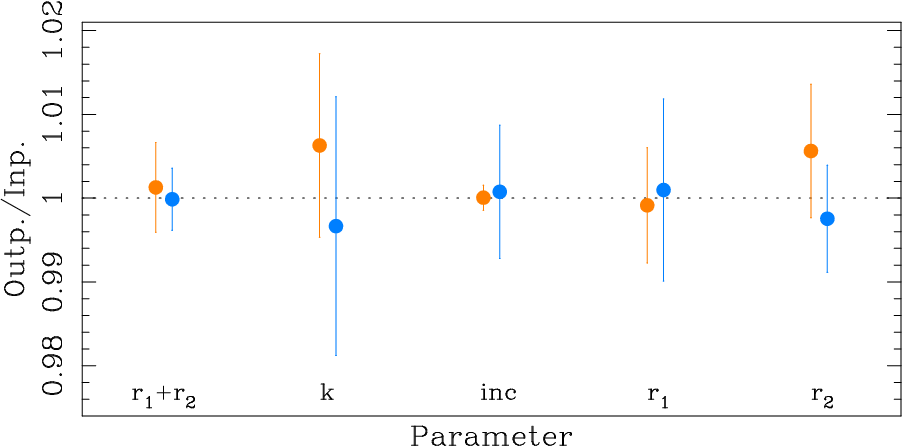}
    \includegraphics[height=4.2cm]{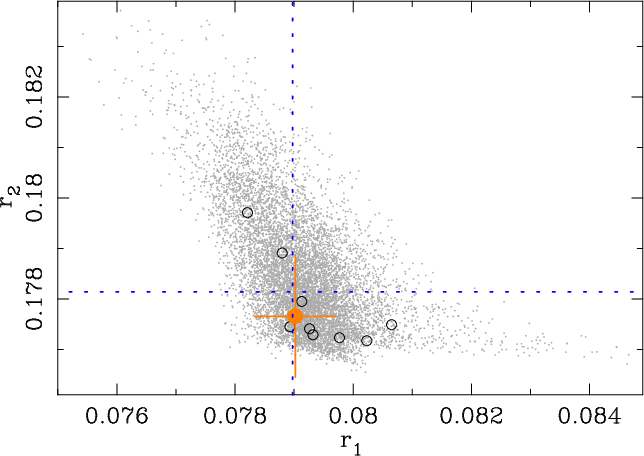}
    \caption{Left: Comparison of results of the \jkt analysis of synthetic ``spotted'' light curves with the
    input values, for a sample of crucial parameters (fractional radii and inclination). The final
    results from all curves/subsets for the binary based on A-045 are plotted with orange symbols, 
    and for the RU~Cnc-based system with blue. In all cases \jkt managed to accurately and precisely
    reproduce the input values.
    Right: Detailed results for the RU~Cnc-based binary, on the $r_1 - r_2$ plane. Grey dots show values 
    from individual MC iterations ($\sim$9000), black circles mark the best-fitting values from nine subsets,
    the orange symbol is the adopted result (with errors), and the blue dashed lines mark the input values.}
    \label{fig_synth}
\end{figure*}

The approach to model a non-trivial light curve of a DEB, by applying additional sines and polynomials, 
has been tested with synthetic data. We used the latest version (v2.3) of the PHysics Of Eclipsing BinariEs 
\citep[PHOEBE;][]{prs05,con20} software to generate eclipsing binary systems with spots that change in time.
We based our synthetic binaries on A-045 and A-083 = RU~Cnc, which have the lowest
average precision in radii in our sample, providing the code with input stellar
parameters (i.e. $M$, $R$, $T_\mathrm{eff}$, $P$, $T_0$) from Table~\ref{tab_par}.
We also used exactly the same times of observations as in real \ktwo curves for these two systems
(from sectors 13 and 5, respectively).
Various values of spot parameters (location, size, relative temperature) were used in each synthetic binary,
so we could better revise their influence on the resulting parameters. In order to mimic the dynamical
character of the synthetic light curve, we assumed non-synchronous rotation of components, and 
synchronicity parameters $F$ were varied from case to case. In this way we made sure that after every
orbital cycle the location of spots changes (migration in longitude) and the shape of the light curve 
is different. When spots were included on different components
we used different values of $F$ for each star. Finally, we added white noise, and (in some cases) additional 
long-term modulation with a 4-th degree polynomial. A total of six different synthetic light curves were
made (three per binary), each showing a different pattern of spots that evolves in time. 

They were analysed with \jkt in the same way as described in Section~\ref{sec_lcfit}, i.e. each was 
divided into subsets, and for each subset we applied a number of sine functions and polynomials.
For each subset, errors were estimated with a Monte-Carlo procedure (task 8).
Results from each individual subset were then weight-averaged, and a combination of their $rms$ and
average error was used as the adopted uncertainty. The resulted values (with errors)
of sum ($r_1+r_2$) and ratio ($k$) of fractional radii, inclination ($i$), and fractional radii
separately ($r_1$ and $r_2$) are compared to their input values in Figure~\ref{fig_synth}.
One can see that the results reproduce the input values with very good agreement, always within 
1$\sigma$. It is worth to note, that while the shape of stars in PHOEBE is dictated by the Roche
geometry, in \jkt they are approximated by 2-axial ellipsoids. This did not affect the results, as 
expected for components with low oblateness. 


\bsp	
\label{lastpage}
\end{document}